\input{psfig.sty}
\documentclass[useAMS]{mn2e}
\usepackage{epsfig}
\usepackage{color}
\usepackage{subfigure}
\usepackage{psfig}
\usepackage{graphicx}
\usepackage{amssymb}

\newcommand{\apj}{ApJ}

\newcommand{\mnras}{MNRAS}

\newcommand{\aap}{A\&A}

\newcommand{\goodgap}{\hspace{\subfigtopskip} \hspace{\subfigbottomskip}}

\title[On the Density profile slope of Clusters of Galaxies]{On the Density profile slope of Clusters of Galaxies}
\author[A. Del Popolo]{A. Del Popolo$^{1,2}$\thanks{E-mail:
antonino.delpopolo@unibg.it} \\
$^{1}$Dipartimento di Fisica e Astronomia, Universit\'a di Catania, Viale Andrea Doria 6, 95125 Catania, Italy\\
$^{2}$ Departamento de Astronomia, Universidade de S\~ao Paulo, Rua do Mat\~ao 1226, 05508-900, S\~ao Paulo, SP, Brazil
\\
}
\begin{document}

%\date{Accepted 1988 December 15. Received 1988 December 14; in original form 1988 October 11}

\pagerange{\pageref{firstpage}--\pageref{lastpage}} \pubyear{2002}

\maketitle

\label{firstpage}    

\begin{abstract}

The present paper extends to clusters of galaxies the study of Del Popolo (2012), concerning  how the baryon-dark matter (DM) interplay shapes the density profile of dwarf galaxies. Cluster density profiles are determined taking into account  dynamical friction, random and ordered angular momentum and  the response of dark matter halos to condensation of baryons. We find that halos containing only DM are characterized by Einasto's profiles, and that the profile flattens with increasing content of baryons, and increasing values of random angular momentum. The analytical results obtained in the first part of the paper were applied to well studied clusters whose inner profiles have slopes flatter than NFW predictions (A611, A383) or are characterized  by profiles in agreement with the NFW model (MACS J1423.8+2404, RXJ1133). By using independently-measured baryonic fraction, a typical spin parameter value $\lambda \simeq 0.03$, and adjusting the random angular momentum, we re-obtain the mass and density profiles of the quoted clusters. Finally, we show that the baryonic mass inside $\simeq 10$ kpc, $M_{b,in}$ is correlated with the total mass of the clusters, 
%finding a correlation among the two quantities, 
as $M_{b,in} \propto M_{500}^{0.4}$.
\end{abstract}


\section{Introduction}

The $\Lambda$CDM model is remarkably successful in fitting a wide range of data of large-scale structures, starting from the anisotropy and polarization spectrum of cosmic microwave background radiation (Spergel et al. 2003; Komatsu et al. 2011), passing through the high-redshift studies of the Ly$\alpha$ forest, and going on with the Hubble diagram of Type Ia Supernovae (Kowalski et al 2008), and the matter power spectrum with its Baryonic Acoustic Oscillation features (Percival et al. 2010).

%Notwithstanding 
In spite of the success of $\Lambda$CDM predictions on cosmological scales, the model has shown difficulties in giving correct predictions on galaxy scales, especially for what concerns the central density 
expected in dark matter haloes. Dwarfs and low surface brightness (LSB) galaxies, objects that are dark matter dominated, are characterized by kinematics, and rotation curves, incompatible with haloes predicted by $\Lambda$CDM, as obtained in dissipationless N-body simulations (Navarro et al. 1996, 1997, hereafter NFW; Moore et al. 1998; Jing \& Suto 2000, Klypin et al. 2001; Hayashi et al. 2004; Power et al. 2003; Navarro et al. 2004, 2010). 
~\\

While dissipationless N-body simulations predict cuspy profiles\footnote{The inner density profile predicted by simulations is characterized by $\rho \propto r^{-1}$ for the NFW model, $\rho \propto r^{-1.5}$ in the Moore et al. (1998) model, and by the Einasto profile, shallowing towards the center of the halo, in more recent simulations (e.g., Navarro et al. 2010). Stadel et al. (2009) found $\alpha \simeq 0.8$ at 120 pc.}, 
observations show that the inner part of density profiles is characterized by a core-like structure (e.g., Flores \& Primak 1994; Moore 1994;  Kravtsov et al. 1998; de Blok \& Bosma 2002; de Blok, Bosma \& McGaugh 2003; Gentile et al. 2004, 2006; Blaise-Ouelette et al. 2004,  Span\'o et al. 2008, Kuzio de Naray et al. 2008, 2009, and Oh et al. 2010). 
The quoted discrepancy between observations and simulations is known as the Cusp-Core problem\footnote{
For the sake of precision, we have to report that some observations found density profiles compatible with both cuspy and cored profiles (van den Bosch et al. 2000; Swaters et al. 2003a,b; Spekkens et al. 2005; Simon et al. 2005; de Blok et al. 2008). }.

Another important test of the results of numerical simulations is to check their predictions also on cluster scales. Constraints on DM density profiles are more straightforward to obtain in clusters of galaxies than in galaxies,\footnote{Even if a drawback in using clusters is that the mass distribution in clusters is usually affected by the growth of the central galaxy.} since clusters have several measurable properties that can be understood and interpreted in a simpler fashion than galaxies rotation curves. For example, X-ray emission from the intracluster plasma at radii of the order of 10\% of the virial radius ($r_{vir}$) can easily observed and the gas temperature profile can be measured with observations (e.g., with Chandra or XMM). 
~\\

Nevertheless many studies have used X-ray observations of the hot intracluster medium, under the assumption of hydrostatic equilibrium, X-ray data alone have difficulties in constraining the mass distribution, especially in the central regions. This because even relaxed clusters tend to have ``cooling flows". In these clusters X-ray emission is often disturbed and the assumption of hydrostatic equilibrium is questionable
%remains an important question 
(see Arabadjis, Bautz \& Arabadjis 2004).
~\\

X-ray temperature measurements are carried out from 500 kpc (Brad$\check{a}$c et al. 2008) to $\simeq 50$ kpc, and the determination of temperature at smaller radii are limited by instrumental resolution or substructure (Schmidt \& Allen 2007). X-ray analyses have obtained wide ranging values of the inner slope, $\alpha$, of the density profile with $\alpha$ ranging from $\simeq$ 0.6 (Ettori et al. 2002) through $\simeq$1.2 (Lewis, Buote \& Stocke
2003) to $\simeq$1.9 (Arabadjis, Bautz \& Garmire 2002), while {\it Chandra} and {\it XMM-Newton} results suggest good agreement with CDM predictions (see Schmidt \& Allen 2007, and references therein).  
The value of $\alpha$ obtained by using X-ray observations have the drawback that it is complicated to take account of the stellar mass contained in the brightest cluster galaxy (BCG), located in the cluster center. The central mass, constituted by stars and gas, even if it is small, compared to the total mass of the system, usually dominates the mass at small radii ($\simeq 10$ kpc), with strong implications on the shape of the inner DM profile.
% (it can mimic a cuspy DM halo if it is not taken into proper account). 
~\\

The dark matter distribution in clusters can be also studied through gravitational lensing. 
Weak lensing of background galaxies is used to reconstruct the mass distribution in the outer
parts of clusters (Mellier 1999). This technique is based on averaging noisy signal coming from many background galaxies. The resolution that can be achieved is able to constrain profiles inside $\sim$100 kpc. An example of its application was given by Dahle, Hannestad \& Sommer-Larsen (2003), who found that the average profile of 6 clusters
%, by them studied, 
agrees with the NFW profile at radii $r \geq 0.1 r_{vir}$, even though with large uncertainties. 
~\\

In the central parts of the cluster, lensing effects become non-linear and one can use the strong lensing technique, in order to constrain the mass distribution.
This technique has a typical sensitivity to the projected mass distribution inside $\sim$100--200 kpc, with limits
%on 
at $\sim$10-20 kpc 
%\citep[e.g.,][]
(Gavazzi 2005; Limousin et al. 2008). 
Typical structures observed in the strong lensing regime are radial arcs, located in positions corresponding to the local derivative of the cluster mass density profile, and tangential arcs whose position is determined by the projected mass density interior to the arc. 
~\\

Example of application of the lensing method are Smith et al. (2001), who studied the tangential and radial arcs of A383, finding an inner slope of the total mass density profile $\alpha>1$ at a radius $\sim 1\% r_{vir}$. Sand et al. (2004), and Newnman et al. (2011), after removing
%when they removed 
the baryonic component, with the aid of stellar kinematics, found that the dark matter only inner slope is flatter than 1. 
%found a flat core, for the DM density profile). 
Kneib et al. (2003) studied Cl 0024+1654 by means of strong and weak lensing finding that a NFW profile fits the profile from $0.1 r_{vir}$ to several values of $r_{vir}$ (but Tyson, Kochansky \& Dell'Antonio 1998 found $\alpha=0.57\pm 0.024$).
%\footnote{Note that Tyson, Kochanski, \& dell'Antonio 1998, found a soft core in the inner profile of the cluster, and the total mass with a slope $\alpha=0.57 \pm 0.02$}. 
Gavazzi et al. (2003), and Gavazzi (2005) studied MS2137.3-2353 concluding that the precise value of the slope depends on the mass-to-light ratio of the BCG (but Sand et al. 2002, found a cored profile with $\alpha \simeq 0.35$).
%\footnote{Sand et al. 2002 found a cored profile}.
~\\

As it is clear from the previous examples, strong gravitational lensing has given controversial results, and in several cases $\alpha$ values are much smaller than $\Lambda$CDM predictions. 
%($\alpha=0.57 \pm 0.024$ for CL 002411654 (Tyson, Kochanski \& Dell'Antonio 1998)\footnote{Kneib et al. 2003, found that a NFW profile fits well the profile}, 
%$\alpha \simeq 0.35$ for MS 2137-23.\footnote{Sand et al. (2002) found that a NFW profile fits well the profile}
%
%is fitted by a NFW profile "although their data seemed to favour an isothermal profile in this
%region".
Moreover, similarly to X-ray analyses, lensing alone cannot disentangle the DM and baryonic components. 
Better constraints on the central part of the density profiles can be obtained through stellar kinematics of the central galaxy ($\sim$1-200 kpc region), and even better results are obtained combining all the previous methods 
%\citep[e.g,][]
(Miralda-Escude 1995; Kneib et al. 2003, (weak+strong lensing); Brada{\v c} et al. 2005 (weak+strong lensing); Mahdavi et al. 2007 (X-ray+weak lensing)). 
%can one hope to achieve a precise and accurate determination of the inner slope. 
~\\

As previously reported, 
%Nevertheless, the different techniques used in the past decade, 
distinct researchers, using different techniques, found results going from good agreement with the $\Lambda$CDM model to disagreement with it, and what is worse is that this often happened for the very same clusters  
%\citep[e.g.,][]
(Smith et al. 2005; Gavazzi et al. 2005; Zappacosta et al. 2006; Schmidt \& Allen 2007; Umetsu \& Broadhurst 2008; Brada{\v c} et al. 2008; Limousin et al. 2008).

The quoted discrepancies in observational results are probably connected to several factors: a) use of observational 
techniques with different/limited dynamic range in radius; b) different definition of the slope, which sometime refers to the DM and sometime to the total mass. We should also not forget the scatter in the profile from cluster to cluster, also observed in the case of dwarfs by Simon et al. (2005), which contributes to the quoted discrepancy between observational results; c) not taking into account the stellar mass of the BCG, which is correct for the outer part of the cluster but wrong for the inner part, where stars dominate the density profile. Concerning this last issue, if the baryon-DM interaction changes the density profile, the usual parameterization of the density profile with NFW or generalized Navarro-Frenk-White (gNFW) models, may be inappropriate (Sand et al. 2008). Another important factor that can produce errors is originated if the clusters shape is not taken correctly into account (Gavazzi 2005). 
%
%%Most clusters, like elliptical galaxies, have a non-spherical shape which is not due to rotation (Rood et al. 1972; Gregory \& Tifft 1976; Dressler 1981). Their %%non-spherical shape is due to the aspherical shape of perturbations which gave rise to them (Barrow \& Silk 1981; Peacock \& Heavens 1985; Bardeen et al. 1986), and by %%further amplification during gravitational collapse (Lin et al. 1965; Icke 1973; Barrow \& Silk 1981). Their shape is supported by velocity anisotropy of galaxies (Aarseth %%\& Binney 1978). Binney \& Silk (1979) and Salvador-Sole \& Solanes (1993) found that the elongation of clusters originates in the tidal distortion by neighboring %%proto-clusters. Inferring mass from kinematic data by assuming spherical symmetry when the actual mass distribution is not spherical induces an error (Gavazzi 2005).
%%If the halo is prolate the dynamical mass is \emph{over}estimated (amplifying the discrepancy we find with NFW), while an
%%oblate halo with short axis along the line of sight causes an downward bias in dynamical mass, which is $\lesssim30\%$ for the above limit on $c$.
%
~\\

A series of papers by Sand (Sand et al. 2002, 2004, 2008, hereafter S02, S04, and S08), further fed the debate on the shape of DM halos of clusters, and the inner slope of
the density profiles. 
%In a series of papers (Sand et al. 2002, 2004, 2008), 
In the quoted papers, Sand combined gravitational lensing with the velocity dispersion profile of the BCG. In this way the author was able to separate the contribution to the halo coming from the DM from that coming from the stellar mass of the BCG. 

S02 studied MS 2137-23 finding a flat inner slope with $\alpha<0.9$ at 99\% CL, and in their best fit parameters, $\alpha$ was 0.35. 

S04 studied a sample of 6 clusters (MS 2137-23; A383; A963; RXJ1133; MACS 1206; A1201), three having just tangential arcs, and three also having radial arcs.
Combining lensing and velocity dispersion profile of the central galaxy in each case, they found a mean DM distribution inconsistent with the NFW value, $\alpha=1$ at 
$>99 \%$ CL. The system of clusters containing radial arcs gives a value of $\alpha=0.52^{+0.05}_{-0.05}$ (68\% CL), and $\alpha <0.57$ at 99\% CL, for the case of the tangential arc sample.   
If their results are correct, they disagree with $\Lambda$CDM model on small scales, unless some physical mechanism not taken into account could get the $\Lambda$CDM out of the trouble (e.g., a form of interplay between DM and baryons). 
Bartelmann \& Meneghetti (2004) and Meneghetti et al. (2007) concluded that the flat slopes obtained by Sand were due to an oversimplified assumptions in the analysis, such as negligible ellipticity.
~\\

An improved analysis of A383 and MS2137-23, presented by S08, found a flat value for the slope of A383 ($\alpha=0.45^{+0.2}_{-0.2}$). Newmann et al. (2009) (hereafter N09), presented a detailed analysis of DM and baryonic distribution in A611, combining weak lensing, strong lensing and stellar velocity dispersion for the BCG, finding a slope $\alpha<0.3$ (68\% CL) and similarly, Newman et al. (2011) (hereafter N11) found $\alpha<1$ (95\% CL) for A383. They showed that degeneracies in constraining the DM profile can be broken only simultaneously using the three techniques. 
The quoted results show that at least some clusters of galaxies have inner density profile slopes shallower that those obtained in N-body 
simulations. 
~\\

These results are in agreement with studies on dwarf galaxies, showing that not all dwarfs have core-like rotation curves and that their density profiles are compatible with both cuspy and core profiles (Hayashi et al. 2004; van den Bosch et al. 2000; Swaters et al. 2003a,b; Simon et al. 2005, Spekkens, Giovanelli \& Haynes 2005; de Blok et al. 2008). 
Very interesting is the case studied by Simon et al. (2003, 2005) who found that in a sample of dwarf galaxies, namely  NGC 2976, NGC 4605, NGC 5949, NGC 5963, and NGC 6689, the inner slopes goe from very flat to cuspy, more precisely $\alpha=(0.01; 0.78; 0.88; 1.20; 0.79)$, respectively, with a mean slope $\alpha \simeq 0.73$, and a dispersion of 0.44. 
~\\

de Blok et al. (2008), using a sample from the HI nearby galaxy survey (THINGS)
found that the best fit to rotation curves depends on their mass: for galaxies having $M_B < -19$ the NFW profile or the pseudo-isothermal profile statistically fitted equally well, while this what not the case galaxies with $M_B > -19$ the core-dominated pseudo-isothermal model fitted significantly better than the NFW model. 
%
%%This argue against universality of density profiles (Subramanian 2000; Ricotti 2003, 2004, 2007; Cen et al. 2004; Simon et al. 2005; Merrit et al. 2005; Graham et al. %%2006; Schmidt et al. 2008; Del Popolo 2009 (hereafter DP09); Ma et al. 2009; Host \& Hansen 2010; Del Popolo 2012 (hereafterDP12)). 
%%It is then interesting to understand if there is a cusp-core problem also for cluster of galaxies, and, as in DP09, to study the effects of different random angular
%%momenta and baryon fraction on the density profile of a given halo. 
%
%
%%In order to save $\Lambda$CDM model and solve the quoted riddle, several studies have shown that one has to take into account the baryon physics and stellar processes in %%the inner parts of galaxies (El-Zant et al. 2001, 2004; Weinberg \& Katz 2002; Romano-Diaz et al. 2008, 2009; Mashchenko et al. 2006; Governato et al. 2010). 
%
%{\bf Aggiustare One of the drawbacks imputed to baryons, whose presence give rise to the adiabatic contraction of dark matter with the result of further steepening the %profile, has been discussed by Zappacosta et al. (2006). X-ray mass measurements in A2589 lead them to conclude that during cluster formation there are processes able to %counteract adiabatic contraction. }
%%In any case, what is certain is that it is needed more observational and theoretical work to understand deeply the baryons-DM interplay in clusters. 
~\\

In several previous papers, we studied the role of baryons on the previously quoted problem (e.g., DP09, DP12).

In the present paper, we extend the analysis of DP12 from dwarf galaxies to clusters of galaxies, studying the effects of baryons
on the inner DM density profile. 
Similarly to DP12, the present paper has two aims.
The first is to study the role of baryons and random angular momentum in changing clusters of galaxies density profiles. The second is to apply the results of this study to observed clusters, to understand if the model can re-obtain the density profile of some well studied clusters with cuspy, cored and intermediate density profiles.   

This paper is organized as follows. Section 2 summarizes the model used. In Section 3.1, we find how changes in baryonic fraction and angular momentum, influence the density profiles 
In Section 3.2, we compare the results of Section 3.1 with four different clusters of galaxies. Section 4 is devoted to discussions and Section 5 to conclusions.

%So, cluster of galaxies are an important laboratory for the quoted tests, because to them are connected several observational probes spanning a large dynamic range in radius and %density. 

\section{Summary of the model}

%\subsection{}

%\subsection{Summary of Del Popolo 2009}

The density profiles of clusters are obtained using the model discussed in DP09, and DP12.
%\footnote{The same model was used in Cardone  et al. (2011), to study dark matter scaling relations in intermediate redshift early - type galaxies, 
%and in Del Popolo 2010, and 2011, to study the problem of universality of the DM density profiles.}. 
For convenience of the reader, the model is summarized in the following of this section. 

DP09 is an improvement of previous spherical infall models (SIM) presented in literature (Gunn \& Gott 1972; Fillmore \& Goldreich 1984; Bertschinger 1985; Hoffman \& Shaham 1985; Ryden \& Gunn 1987; Avila-Reese, Firmani \& Hernandez 1998; Subramanian et al. 2000; Ascasibar, Yepes \& G\"ottleber 2004; Williams, Babul \& Dalcanton 2004). 
~\\

Differently from previous SIMs, DP09 includes in the model the joint effects of ordered and random angular momentum, dynamical friction, adiabatic contraction of dark matter.
%, an baryons-DM interplay. 
Previous SIMs usually took account one effect at a time: random angular momentum (e.g., Williams, Babul \& Dalcanton 2004), dynamical friction of stellar/DM clumps against the background halo (e.g., El-Zant et al. 2001; Romano-Diaz et al. 2008), or just adiabatic contraction (e.g., Blumenthal et al. 1986; Gnedin et al. 2004; Klypin,  Zhao, and Somerville 2002; Gustafsson et al. 2006). 
~\\

The main features of the SIM are described by Gunn \& Gott (1972). 
The quoted model assumes that a protostructure is divided into mass shells, each one expanding with the Hubble flow from an initial comoving radius $x_i$ to a maximum one $x_m$ (usually named turn-around radius $x_{ta}$), and then collapse. Non-linear processes convert the kinetic energy of collapse into random motions, giving rise to a ``virialized" structure. 
The density profile at turn-around, $\rho_{ta}(x_m)$, is obtained following the dynamics of collapse of the shells, assuming that mass is conserved and that each shell is kept at its turn-around radius (Peebles 1980; Hoffman \& Shaham 1985; White \& Zaritsky 1992). 
~\\

In order to go on from turn-around to the final collapse, and to obtain the final density profile, one usually assumes that the potential well near the center varies adiabatically (Gunn 1977, Fillmore \& Goldreich 1984). The final density is given by 
\begin{equation}
\rho(x)=\frac{\rho_{ta}(x_m)}{f^3} \left[1+\frac{d \ln f}{d \ln x_m} \right]^{-1}
\label{eq:dturnnn}
\end{equation}
where $f=x/x_m$ is the collapse factor (see Eq. A18, DP09). 
During expansion, the tidal fields originated by the large-scale structure exert a torque on the proto-structure imparting angular momentum on the proto-halo (Hoyle 1953; Peebles 1969; White 1984; Ryden 1988; Eisenstein \& Loeb 1995; Catelan \& Theuns 1996). After that the proto-strucure decouples from the background, turns-around and collapses, tidal torque is noteworthy reduced, since the lever arm is reduced drastically by collapse (Schaefer 2009). 
Angular momentum originated by tidal torques is calculated obtaining the rms torque, $\tau (r)$, on a mass shell  and then 
calculating the total specific angular momentum, $h(r,\nu )$, acquired during expansion by integrating the torque over time (Ryden 1988, Eq. 35).
As shown in DP09, the values obtained for angular momentum are in agreement with others studies (e.g. Catelan \& Theuns 1996).
As in DP12, total angular momentum is expressed in terms of 
%of a system is often expressed in terms of 
the dimensionless spin parameter
\begin{equation}
\lambda=\frac{L |E|^{1/2}}{GM^{5/2}}, 
\end{equation}
where $L$ is the angular momentum, summed over shells, and $E$ is the binding energy of the halo. 
According to Padmannabhan (1993), $\lambda$ can be interpreted as the ratio of the angular velocity, $\omega$ of the system to that, $\omega_{sup}$ of the system providing the rotational support, obtaining
\begin{equation}
\lambda=\frac{\omega}{\omega_{sup}}=\frac{L}{2G^{1/2}M^{3/2} R^{1/2}}, 
\end{equation}
In a system constituted by DM and baryons as ours, we have a spin parameter for DM and another for gas
\begin{equation}
\lambda_{gas(DM)}=\frac{L_{gas(DM)}}{M_{gas(DM)} [2G(M_{gas}+M_{DM})r_{vir}^{1/2}]}, 
\end{equation}
where $L_{gas(DM)}$ is the angular momentum of gas(DM), and $M_{gas(DM)}$ the gas(DM) mass inside the virial radius $r_{vir}$.
The ratio of $\lambda_{gas}/\lambda_{DM}$ is fixed according to G\"ottleber \& Yepes (2007) (1.23 for haloes with $M_{vir}> 5 \times 10^{14} h^{-1} M_{\odot}$ (see their figure 5)). 
The distribution of the $\lambda$ parameter is well described by a lognormal distribution (e.g. Vivitska et al. 2002).
%Using the parameters in Vivitska et al. (2002), the maximum of the distribution of $\lambda$ is $\lambda=0.035$, while there is a 90\% probability that $\lambda$ is in the %range 0.02-0.1. 
G\"ottleber \& Yepes (2007) studied the spin parameter of 10,000 clusters extracted from the {\it Mare Nostrum Universe} SPH simulation, finding a log-normal distribution with best fit parameters given by  $\lambda$ with $\lambda=0.0351 \pm 0.0016$, $\sigma_{\lambda}=0.6470 \pm 0.0067$ for the DM distribution and $\lambda=0.0462 \pm 0.0012$, $\sigma_{\lambda}=0.6086 \pm 0.0030$ for the gas distribution, with $\lambda_{\max}=0.0231, 0.0319$ for DM and gas distributions (see also Bett et al. 2007; Sharma \& Steinmetz 2005).
A spin parameter of $\lambda \simeq 0.05$ is characteristic of objects having negligible rotational support, and little systematic rotation (e.g., large elliptical galaxies and clusters of galaxies). Spiral and LSB galaxies are rotationally supported and in the case of LSBs $0.06<\lambda<0.21$ are LSBs (Jimenez et al. 1998; Boissier et al. 2003). 

%%{\bf
Apart from this source of angular momentum, originated by bulk motions and tidal fields, a random angular momentum, $j$, is generated by random velocities (see Ryden \& Gunn 1987).  
%and the following of the present paper), 
In several of the papers which took account of angular momentum in SIM (e.g., Nusser 2001, Hiotelis 2002; Williams et al. 2004) only random angular momentum 
%the second type of angular momentum (the one generated by random motions) 
was taken into consideration. The approach usually followed was that of assigning a specific angular momentum at turn around (e.g., Nusser 2001; Hiotelis 2002; Ascasibar, Yepes \& G\"ottleber 2004), having typical value:
\begin{equation}
j=j_{\ast} \propto \sqrt{GM r_m}  
\end{equation}
The specific angular momentum, $j$, may be also expressed through the ratio $e_0=\left( \frac{r_{min}}{r_{max}} \right)_0$, where $r_{min}$ and $r_{max}$, are the pericentric and apocentric radii.
% (Avila-Reese et al. 1998), which is usually left as a free parameter. 
Avila-Reese et al. (1998) used $e_0$ as a free parameter to take into consideration processes related to mergers and tidal forces that could produce tangential perturbations. In their paper, they showed that the detailed description of these processes is largely erased by the virialization process, remaining only through the value of $e_0$, which they fixed to $e_0=0.3$. The value $e \simeq 0.2$ gives density profiles very close to the NFW profile (Avila-Reese et al. 1998, 1999). The previous procedure is based on results of N-body simulations of CDM halo collapse, giving constant 
$< \frac{r_{min}}{r_{max}}> \simeq 0.2$ ratios of dark matter particles in virialized haloes.
An improvement to the previous model is due to Ascasibar, Yepes \& G\"ottleber (2004), who found that particle orbits are slightly more radial as we move out to the current turn around radius, $r_{ta}$, and that there is a dependence on the dynamical state: major mergers are well described by constant eccentricity up to the virial radius, relaxed systems are more consistent with a power-law profile, while minor mergers are in the middle.   
%The average profile can be fitted by a power law, but the slope is shallower than for relaxed systems. 
A least-square fit to the relaxed population yields:
\begin{equation}
e(r_{max}) \simeq 0.8 (r_{max}/r_{ta})^{0.1}
\end{equation}
for $r_{max}< 0.1 r_{ta}$.

In the present paper, random angular momentum is taken into account through 
%we use 
Avila-Reese et al. (1998) method with the correction of Ascasibar, Yepes \& G\"ottleber (2004).
%It is important to notice that even using a constant value for the eccentricity angular momentum will change with radius and $\nu$.
%In fact the specific angular-eccentricity relation is:
%\begin{equation}
%j = \sqrt{G M r_{max} (1-e)}
%\end{equation}
%which depends from mass (radius) and is anti-correlated with $\nu$ ($r_{m} \propto 1/\nu$). 

%}

Dynamical friction was taken into account as described in Antonuccio-Delogu \& Colafrancesco (1994) (see also Appendix D of DP09). 
%calculated dividing the gravitational field into an average and a random component generated by the clumps constituting hierarchical universes.
Its effects on structure formation were obtained by introducing the dynamical friction force in the equation of motion (Eq. A14 in DP09). 
%{\bf 
We also took into account adiabatic contraction (AC) of dark matter halos in response to the condensation of baryons in their centers, and leading to a steepening of the dark matter density slope. 
%%Adiabatic contraction was taken into account by means of Gnedin et al. (2004) scheme, which is an improvement of that of Blumenthal et al. (1986).  
%%Exchange of angular momentum between baryons and DM (e.g., through dynamical friction), was taken into account through Klypin et al. (2002) model.
%taking also account of exchange of angular momentum between baryons and dark matter. 

We summarize in the following the AC model taking also account of angular momentum exchange. 

The present standard form of the AC model was introduced and tested numerically by Blumenthal et al. (1986) (see also Ryden \& Gunn 1987; Barnes 1987; Oh 1990).  
Let's consider a spherically symmetric protostructure that consists of a fraction of baryons, $F_b=M_b/M_{500} <<1$, and a fraction $1-F_b$ of dark matter particles constituting the halo. Here, $M_{500}$ is the mass contained within a radius where the enclosed density is 500 times the critical density of the Universe, and $M_b=M_{\ast}+M_{gas}$, where $M_{\ast}$ is the stellar mass and $M_{gas}$ is that of gas. 
Baryons and halo particles are assumed to be well mixed initially (i.e., the ratio of their densities is $F_b$ through the protostructure). As the baryons dissipatively cool and fall into a final mass distribution $M_b( r)$, 
%which is constrained by the initial angular momentum distribution, 
a dark matter particle initially at radius $r_i$ will move in to  radius $r< r_i$,
 %The adiabatic invariant for such a particle orbits implies that:
characterized by
\begin{equation}
r \left [ M_b( r) +M_{dm} ( r) \right] = r_i M_i (r_i)
\label{eq:ad1}
\end{equation}
%which can be solved iteratively for the final distribution of dissipationless halo particles 
%$M_{dm} ( r)$ given the initial total mass distribution $M_i (r_i)$ and the final baryon mass distribution $M_b( r)$.
where  $M_i (r_i)$ is the initial total mass distribution, $M_b( r)$ is the final mass distribution of dissipational baryons and $M_{dm}$ is the final distribution of dissipationless halo particles. 
Assuming that the orbits of the halo particles do not cross, one obtains
\begin{equation}
M_ {dm} ( r)=(1-F_b) M_i (r_i)
\label{eq:ad2}
\end{equation}
Eqs. (\ref{eq:ad1}), (\ref{eq:ad2}) can be iteratively solved 
%used 
to calculate the final radial distribution of the halo particles once $M_i (r_i)$ and $M_b (r)$ are given.
%In order to use the adiabatic invariant, one has to know the initial mass distribution $M_i (r_i)$ and the final distribution of dissipational baryons
%$M_b (r)$. 
A usual assumption is that initially baryons had the same density profile as the dark matter (Mo  et al. 1998; Cardone \& Sereno 2005; Treu \& Koopmans 2002; Keeton 2001), 
%that initial dark matter density profile is described by a NFW profile,
and the final baryons distribution is assumed to be 
%a disk (for spiral galaxies) (Blumenthal et al. 1986; Flores et al. 1993; Mo et al. 1998; Klypin et al. 2002; Cardone \& Sereno 2005). In our %calculations, we shall assume Klypin et al (2002) model for baryons distribution (see their subsection 2.1), when dealing with mass scales typical of spiral galaxies. 
a Hernquist configuration (Rix et al. 1997; Keeton 2001; Treu \& Koopmans 2002). In the present paper, the Blumenthal's model was improved following Gnedin et al. (2004) who proposed a simple modification which describes numerical results more accurately.
They proposed a modified adiabatic contraction model based on conservation of the product of the current
radius and the mass enclosed within the orbit-averaged radius:
\begin{equation}
  M(\bar{r})r= {\rm const}.
  \label{eq:modified}
\end{equation}
where the orbit-averaged radius is
\begin{equation}
  \bar{r} = {2 \over T_r} \int_{r_{min}}^{r_{max}} r \, {dr \over v_r},
\end{equation}
where $T_r$ is the radial period.
%, $r_a$ is the apocenter radius and $r_p$ the pericenter radius.  

The previous, AC model, assumes no angular momentum exchange between different components (e.g., baryons
and DM). In the early phase of collapse, the baryons density is an order of magnitude smaller than that of DM, and the exchange of angular momentum has little impact on the dark matter. During the collapse baryons infall into the center of the proto-structure deepening the gravitational potential well.
In the late stage of collapse the baryon density becomes large, non-axisymmetric component may develop due to the excitation of spiral waves and/or bar-like modes.
%, and the increase in DM density acts as a sort of coupling process of the baryons and the DM (Klypin et al. 2001; Klypin,  Zhao, and Somerville 2002).
%In this case, the exchange happens at late stage of baryonic infall when the baryon density becomes large and a non-axisymmetric
%component may develop due to the excitation of spiral waves and/or bar-like modes. 
The resulting increase of DM density acts as a sort of coupling process of the baryons and the DM (Klypin et al. 2001; Klypin,  Zhao, and Somerville 2002). 
%Then itself AC provides a coupling between the baryons and the dark matter.
At later stages of collapse, when baryon density increases, the effect is potentially quite large resulting in a decrease of the dark matter density by a factor of ten (Klypin, Zhao, \& Somerville 2002). 
%The description of angular momentum transfer through dynamical friction was described in Sect. 2 and more widely in Appendix E of DP09.

In order to describe quantitatively the quoted process, let's consider a spherical shell of dark matter 
%with radius $r$. If one considers a spherical shell of dark matter 
with density $\rho_{\rm dm}$, radius $r$, thickness
$dr$, and specific angular momentum
\begin{equation}
j=r V_c = \sqrt{G\left[M_{\rm b}(r)+M_{\rm dm}(r) \right] r}.
\label{eq:jr}
\end{equation}
%where $M_{\rm b}(r)$ is the mass distribution of baryons, and $M_{\rm dm}(r)$ is the distribution of dissipationless halo particles, 
one can get an implicit equation for the final radius $r_f$:
\begin{eqnarray}
\label{eq:exchangeangularA}
j_f &=& j \left[1+\frac{A\Delta M}
                  {4\pi\rho_{\rm dm}r^3}\right],\phantom{mmm} \\
A &=& 1+\frac{r}{V_c}\frac{dV_c}{dr},\\
\Delta M &=&M_{\rm b,f}-M_{\rm b}.
\label{eq:exchangeangularB}
\end{eqnarray}
Klypin,  Zhao, and Somerville (2002), where $M_{\rm b,f}$ is the final baryons mass and $M=M_{\rm dm}+M_{\rm b}$ is the total mass inside a radius $r$.
Eq.~(\ref{eq:exchangeangularA}) is solved numerically. The solution also gives the mass inside a final radius $r_f$.
Eq.~(\ref{eq:exchangeangularA}) has the same structure as Eq.~(\ref{eq:ad1}), the only difference is the term on the
right-hand-side, which is the correction due to angular momentum deposition.

%}

For what concerns baryons, we discussed how they were introduced in appendix E of DP09, and how their initial mass distribution and the final distribution were fixed. 
%In the case of clusters, 
%{\bf
The quantity of baryons used in the calculation of Sect. 3.1, was fixed using Giodini et al. (2009) results. 
%}
%and McGaugh et al. (2010) studies. 
As shown from McGaugh et al. (2010) (see their Fig. 2), the fraction of baryons detected in all forms deviates monotonically from the cosmic baryon fraction as a function of mass. On the largest scales of clusters, most of the expected baryons are detected, while in the smallest dwarf galaxies, fewer than 1\% are detected. 
The detected baryon fraction, $f_d$, is the ratio of the baryon fraction, $F_b=M_b/M_{500}$, and the universal baryon fraction, $f_b$, is 
\begin{equation}
f_d = (M_b/M_{500})/f_b=F_b/f_b
\end{equation}
where $f_b=0.17 \pm 0.01$ (Komatsu et al. 2009).
Note that in this paper the masses, $M_{200}$, and $M_{vir}$, are converted to $M_{500}$ using the method used in White (2001), Hu \& Kravtsov (2003), and Lukic et al. (2009). 
%The cluster mass $M_{cl}$ is $M_{500}$.
 
In the four clusters
%example??? clusters 
studied in Sect. 3.2, we calculated the
%used the baryon fraction that was estimated by papers devoted to study in detail the quoted clusters (see the following), and also using Schmidt \& Allen (2007) in our example clusters %by them also studied. 
baryon fraction using Schmidt \& Allen (2007) 
%(for example in the case of A611, A383, MACS J1423.8+2404), 
and checked that the result was in agreement with Giodini et al. (2009) results.
%, and McGaugh et al. (2010).
More in detail, in this last case, we calculated the baryon content subtracting the total mass of clusters, obtained from the data in table 3 of Schmidt \& Allen (2007), from the DM mass obtained from the data in their table 4, namely
\begin{equation}
M_b=M_{DM+b}-M_{DM}= \frac{800 \pi}{3} \rho_c (r^3_{vir,DM+b}-r^3_{DM})
\label{eq:mass}
\end{equation}
where $\rho_c$ is the mean background (critical) density, and $r_{vir}=c r_s$, where the concentration parameter $c$, and the scale factor $r_s$ are given in Schmidt \& Allen (2007) for 34 massive, dynamically relaxed clusters of galaxies.
% (including some of the example clusters that we used). 
%The baryon content estimated with these different methods are in agreement.
Note, that Eq. (\ref{eq:mass}) comes from the fact that for a NFW profile $M(r=r_{vir})=\frac{800 \pi}{3} \rho_c r^3_{vir}$.
~\\

%{\bf

In order to have a more clear view of how the previous quoted processes works to build up the cluster, we summarize the structure formation starting from 
high redshifts. 
%More in detail, a description of how dynamical friction, angular momentum, and baryons and DM interplay to give the final density profile is described %in the text relative to Fig. 5 of DP09. 
Initially the proto-structure is in the linear phase, it expands, reaches a maximum of expansion and then collapses. Baryons trapped inside the potential wells of DM halos are subject to radiative dissipation processes which give rise to clumps and self-gravitating clouds before it collapses to the halo center and condenses into stars and galaxies. The stage of baryons cooling and stars formation happens as described in Ryden (1988) (Sect. 4: ``baryionic dissipation"). In the infall, the baryons compress the DM halos (AC),
%, the so called adiabatic contraction (AC), 
producing a steepening of the DM density profile. When clumps reaches the central high density regions
%once the baryons condense to form stars and galaxies, 
they experience a dynamical friction force from the less massive DM particles as they move through the halo. Dynamical friction acts as an angular momentum engine (Tonini, Lapi \& Salucci 2006), and energy and angular momentum are transferred to DM (El-Zant et al. 2001, 2004), increasing its random motion, and giving rise to a motion of DM particles outwards\footnote{ Another model of baryon-DM interaction, and exchange of angular momentum between the baryons and DM was previously discussed in this section (Klypin et al. 2001; Klypin, Zhao, \& Somerville 2002). }.
Moreover, ordered angular momentum mainly acquired in the expansion phase, through tidal torques, gives rise to nonradial motions in the collapse phase that amplifies the effects of the previous mechanism. Then the joint effect of angular momenta (ordered and random) and dynamical friction, transporting it from baryons to DM, overcomes that of the AC and the profile starts to flatten. 

As previosly described, in order to follow the structure formation, the protostructure is divided into mass shells, containing DM and baryons, which after radiative dissipation forms clumps, stars, and galaxies. 
In order to form the structure the evolution in various phases is followed by means of the SIM. After turn-around shells move inward, baryons (namely, baryonic clumps) exchange angular momentum with DM and accretes to the center, forming the central galaxy (BGC) (see also Lackner \& Ostriker 2010).  

%}

%Specifically, in the very early stages of galaxy formation, the baryons trapped inside the potential wells of the halos undergo radiative dissipation %processes that cause them to lose kinetic energy and to form clumps inside the relatively smooth dark halo. If radiative cooling is effective, the gas %will organize into selfgravitating clouds before it collapses to the halo center and condenses into stars; the clouds are likely to survive the tidal %stripping due to the DM, because of their relatively high binding energy (Mo \& Mao 2004). The clumpy gas component decouples from
%virial equilibrium and dissipates its orbital energy; in detail, the clouds spiralling down get closer and closer to the halo center, increasing
%their tangential velocity along their orbits and reaching regions with higher and higher density. At each instant, the force exerted by the %backgroundDMparticles

As already shown in DP09, and DP12, the density profiles, obtained with the model summarized in this section, and fully described in DP09, DP12, produces profiles in agreement with other studies (e.g., El-Zant et al. 2001, 2004; Mashchenko et al. 2005, 2006, 2007; Romano-Diaz et al. 2008, and the more recent paper of Governato et al. 2010).

\section{Results and discussion}

\subsection{Changes of density profiles with baryons fraction and angular momentum}

As discussed in the introduction, dwarf galaxies and LSBs have rotation curves incompatible with the haloes that $\Lambda$ CDM model predicts. This problem, known as Cusp-Core problem, 
pointed out by Flores \& Primack (1994), and Moore (1994) has not seen till now a clear-cut solution. 
%Several studies have shown that to save $\Lambda$CDM model and solve the quoted riddle, one has to take into account the baryon physics and stellar processes in the inner parts of %galaxies. Interactions of the DM with a stellar bar (Weinberg \& Katz 2002; McMillan \& Dehnen 2005), decay of binary black hole orbits after galaxies merge (Milosavljevi\'c \& Merritt %2001), baryon energy feedback from
%active galactic nucleus (Peirani et al. 2008), dynamical friction of stellar/DM clumps against the background DM halo (El-Zant et al. 2001, 2004; Romano-Diaz et al. 2008, 2009), random %bulk motions of gas in primordial galaxies, driven by supernova explosions (Mashchenko et al. 2006), removal of low-angular momentum gas (Governato et al. 2010), are some of the %solutions proposed. BIBLIOGRAFIA IN LACKNER. IL MIO DOV'e'?  

Inner density profiles in clusters of galaxies have also been at the center of a similar debate to that of dwarfs, with claims of value of $\alpha$ going from $\alpha=0.35$ (S02) %through intermediate values 
to $\simeq$1.9 (Arabadjis, Bautz \& Garmire 2002).

To this aim, in the following of the paper, we use DP09 to generate density profiles in the range $10^{14}-10^{15} M_{\odot}$, then focalizing our attention on changes of inner density profiles produced by different values of baryon fraction and random angular momentum. 

In Sect. 3.1, and 3.2 the baryon fraction was obtained as described in the final part of Sect. 2.  
%obtained by Giodini et al. (2009), McGaugh et al. (2010), and Schmidt \& Allen (2007) (see Sect. 2 for an explanation). 
%In Sect. 3.2, we compare the result of the model to some observed clusters, and discuss how baryon fraction is chosen for those peculiar cases. 
Ordered angular momentum is fixed to typical values obtained by Gottl\"ober \& Yepes (2007), and random angular momentum is used as a parameter.  

The results of our calculation are plotted in Fig. 1a-c. 
%
%%In Fig. 1a, the solid line shows the density profiles of haloes obtained in DP09 (which we will use as reference halo), in the case $10^{14} M_{\odot}$.  
%
The ordered angular momentum used to obtain the solid line in Fig. 1a-1c (obtained by the tidal torque theory as in DP09) is characterized by $\lambda=0.03$\footnote{Note that even if we studies haloes of mass in the range $10^{14}-10^{15} M_{\odot}$, and this imply differences in ordered angular momentum, the spin parameter, $\lambda$ has small differences.}. The baryonic fraction is $F_{B_{\ast}}=M_{b}/M_{500}=0.15$, or $f_d=F_b/f_b \simeq 0.88$ (see McGaugh et al. 2010), and the typical random angular momentum is $j_{\ast}$. The solid line in Fig. 1a-c is obtained using these typical values. The solid line, obtained as described is chosen as the reference halo. 

\begin{figure*}
\centering
(a)
\hspace{0.8cm}
\subfigure{\includegraphics[width=7.4cm]{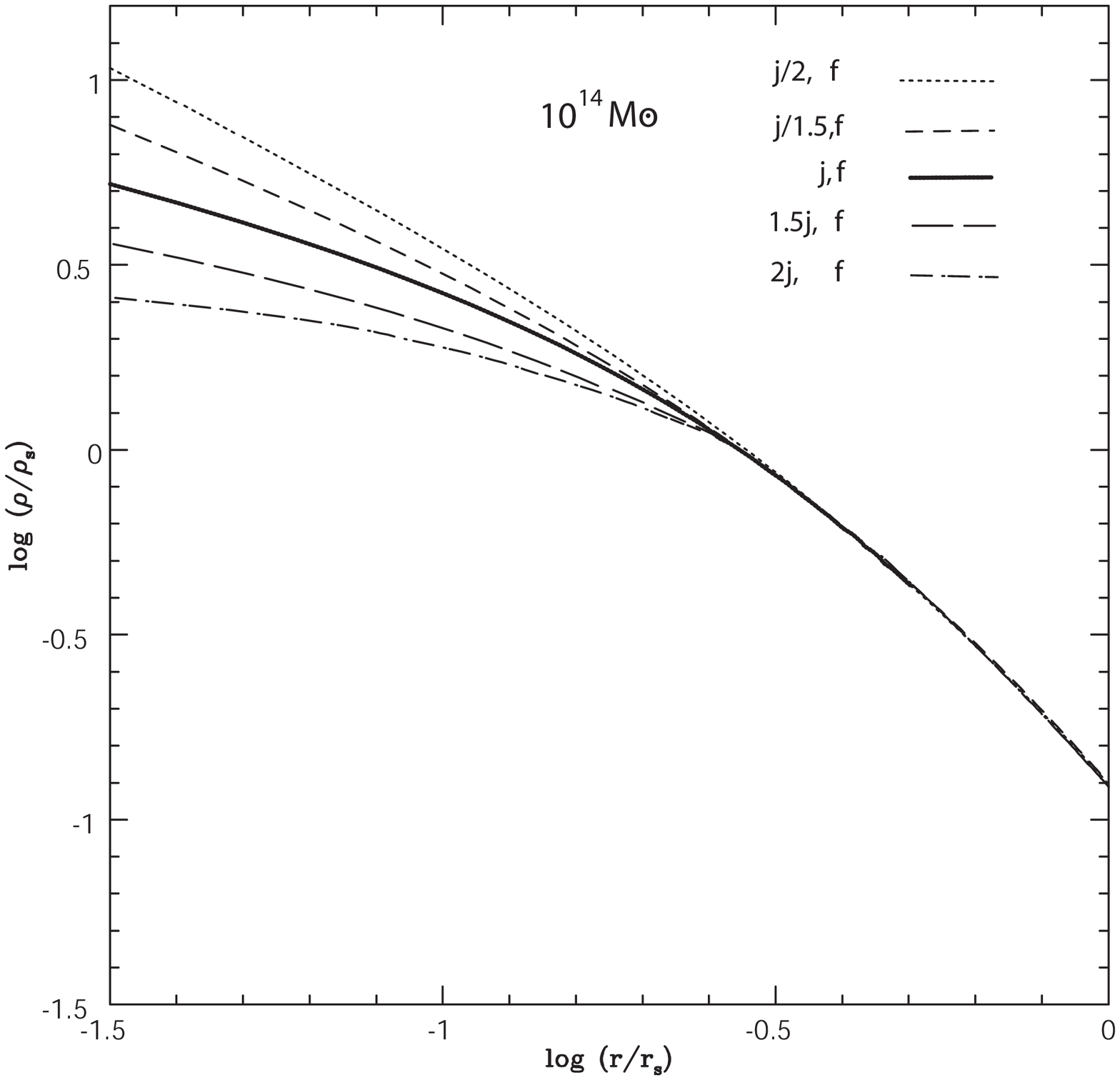}} (b) \goodgap 
%\includegraphics[width=84mm]{dwarfsnmm.eps}
%\hspace{-0.15cm}
\subfigure{\includegraphics[width=7.6cm]{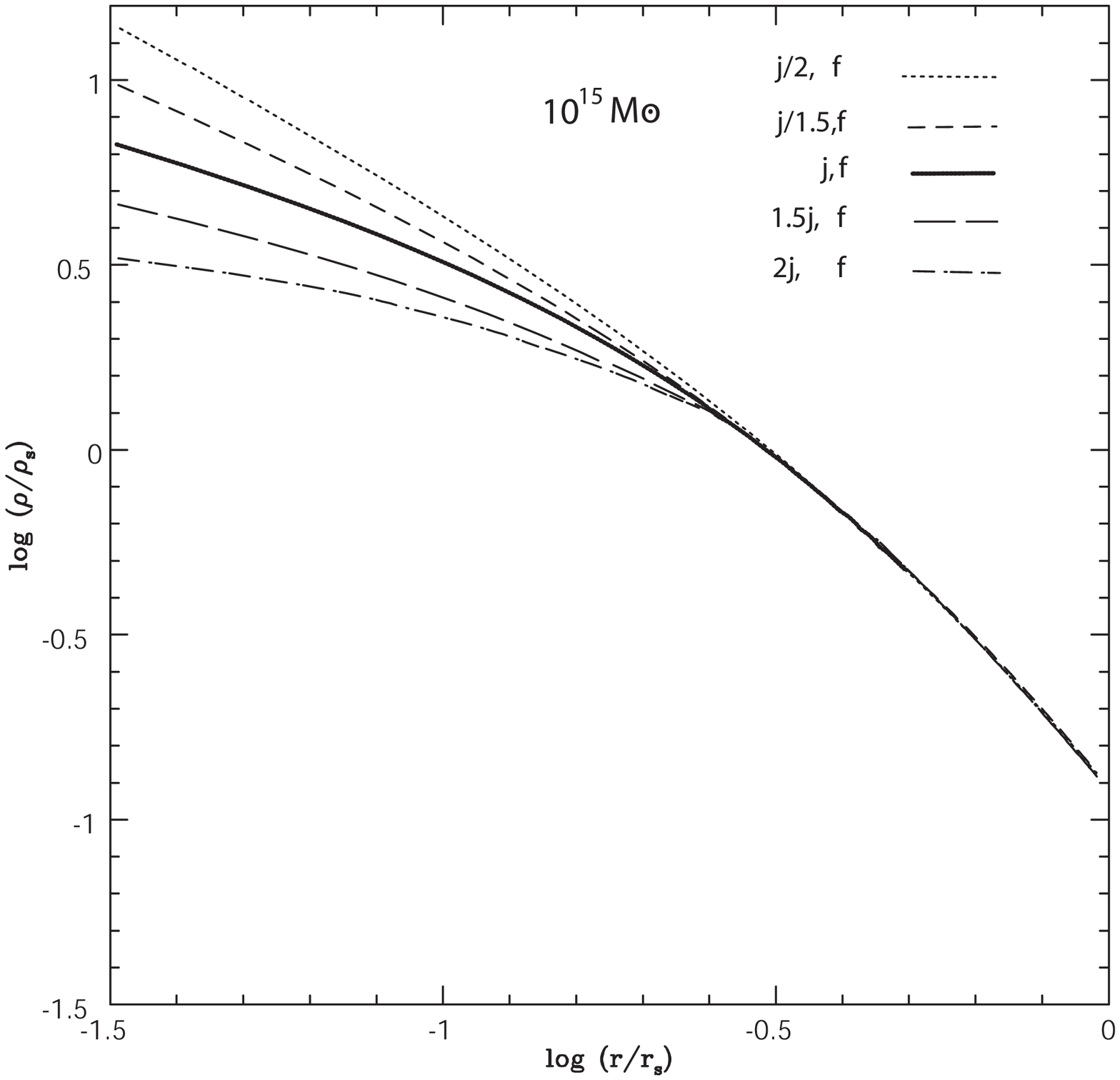}}  \\ (c)\goodgap 
%\hspace{-0.5cm}
\subfigure{\includegraphics[width=7.8cm]{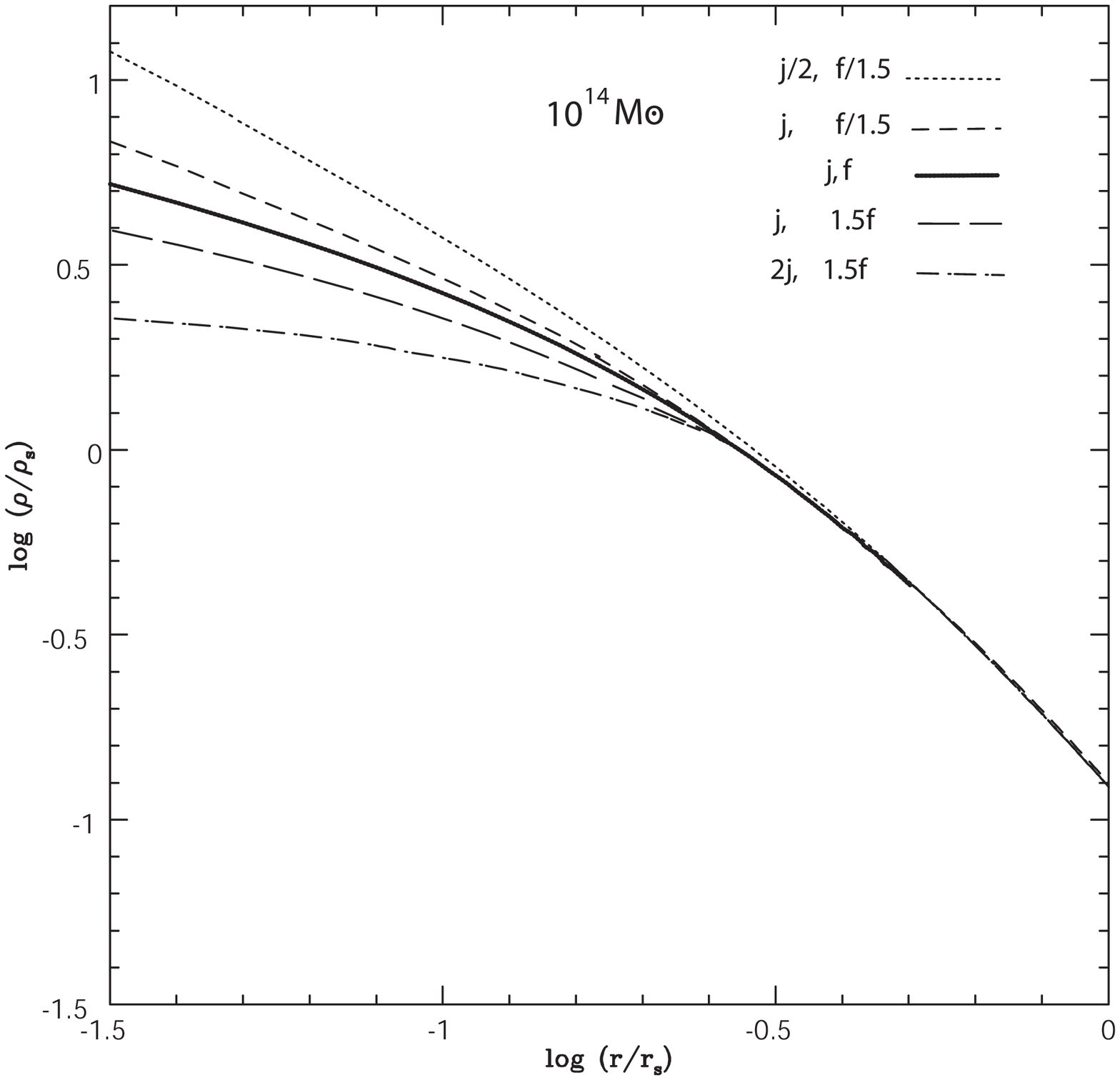}} (d)  \goodgap\\  
%\includegraphics[width=84mm]{dwarfs33.eps}
%\hspace{0.15cm} \includegraphics[width=68mm]{dwarfs4i.eps}
%\hspace{-0.15cm} 
%%%%%%%%%%%%%%%%%%%%%%%%%%%%%%%%%%%%%%%%%%%%%\subfigure{\includegraphics[width=6cm]{einas.ps}} \goodgap \\0
%\hspace{0.15cm} \includegraphics[width=68mm]{dwarfs4ii.eps}
%\includegraphics[width=84mm]{dwarfsn_isol.eps}
\caption{Changes of the density profile with baryonic fraction and random angular momentum $j$. Panel (a) refers to density profiles with mass $10^{14} M_{\odot}$ varying only $j$. The solid line is characterized by 
$F_{B_{\ast}}=M_{b}/M_{500}=0.15$ ($f_{d_{\ast}}=0.88$), and $j=j_{\ast}$. The upper short-dashed line, and the dotted line are characterized by 
$f_d=f_{d_{\ast}}$, and $j=j_ {\ast}/1.5$, and $f_d=f_{d_{\ast}}$, $j=j_{\ast}/2$, respectively. The long dashed line, and dot-dashed line, are characterized by 
$f_d=f_{d_{\ast}}$, and $j=j_ {\ast} \times 1.5$, and $f_d=f_{d_{\ast}}$, $j=j_{\ast} \times 2$, respectively. Panel (b), represents the density profile for haloes of mass $10^{15} M_{\odot}$. Panel (c) plots density profiles for haloes of mass $10^{14} M_{\odot}$ varying both $j$ and baryonic fraction. The solid line, short-dashed line, dotted line, are characterized by $f_d=f_{d_{\ast}}$, $j=j_{\ast}$; $f_d=f_{d_{\ast}}/1.5$, $j=j_{\ast}$; $f_d=f_{d_{\ast}}/1.5$, $j=j_{\ast}/2$, respectively. The long-dashed line, dot-dashed line, are characterized by $f_d=1.5 \times f_{d_{\ast}}$, $j=j_{\ast} $; $f_d=1.5 \times f_{d_{\ast}}$, $j=j_{\ast} \times 2$, respectively.}
\end{figure*}

%Ordered angular momentum is fixed to a typical value ($\lambda \simeq 0.03$). 
%Differently from DP12, we do not consider changes in spin parameter, $\lambda$, since differently from dwarf galaxies, clusters are not rotationally supported and we assume a similar %value of $\lambda$.
Differently from DP12, in the present paper, we will study the change of the inner density profile slope just when the random angular momentum changes, since differently from dwarf and spiral galaxies, clusters of galaxies are non-rotationally supported (the mean rotational velocity, and ordered angular momentum of most luminous clusters are much less than their velocity dispersions).
%This because while dwarf galaxies are rotational supported and ordered angular momentum is of fundamental importance in shaping their density profiles, clusters have %negligible rotational support, and 
%We assume, for the case studied, a similar value of $\lambda$.

In fig. 1a, we fix the value of $f_d$ to the value previously discussed, and vary the amplitude of the random angular momentum, $j$. 
The short-dashed (dotted) line in Fig. 1a, represents the case $j=j_{\ast}/1.5$ ($j=j_{\ast}/2$), where $j_{\ast}$ is the value of $j$ for the reference case (solid line).
%and is equal to...........
As expected, the density profile steepens when $j$ decreases ($\alpha \simeq 0.6$ (solid line), 0.75 (short-dashed line), 0.95 (dotted line)).
%, for the solid, short-dashed, and dotted line, respectively). 
This steepening of the profile with $j$ agrees with several previous results (Sikivie et al. 1997; Avila-Reese et al. 1998; Nusser 2001; Hiotelis 2002; Le Delliou \& Henriksen 2003; Ascasibar, Yepes \& G\"ottleber 2004; Williams et al. 2004). The long-dashed (dot-dashed) line in Fig. 1a, represents the case $j=j_{\ast} \times 1.5$ ($j=j_{\ast} \times 2$). As expected, larger values of $j$ produce flatter profiles ($\alpha \simeq 0.4$, 0.2, for the long-dashed, and dot-dashed line, respectively).  

Fig. 1b, is the same as Fig. 1a, but the halo mass is in this case $10^{15} M_{\odot}$. Short-dashed (dotted) line is characterized by $\alpha \simeq 0.93$ (1.05), while long-dashed (dot-dashed) line is characterized by $\alpha \simeq 0.52$ (0.32). 
\begin{figure*}
\centering
(a)
\hspace{0.8cm}
\subfigure{\includegraphics[width=14.4cm]{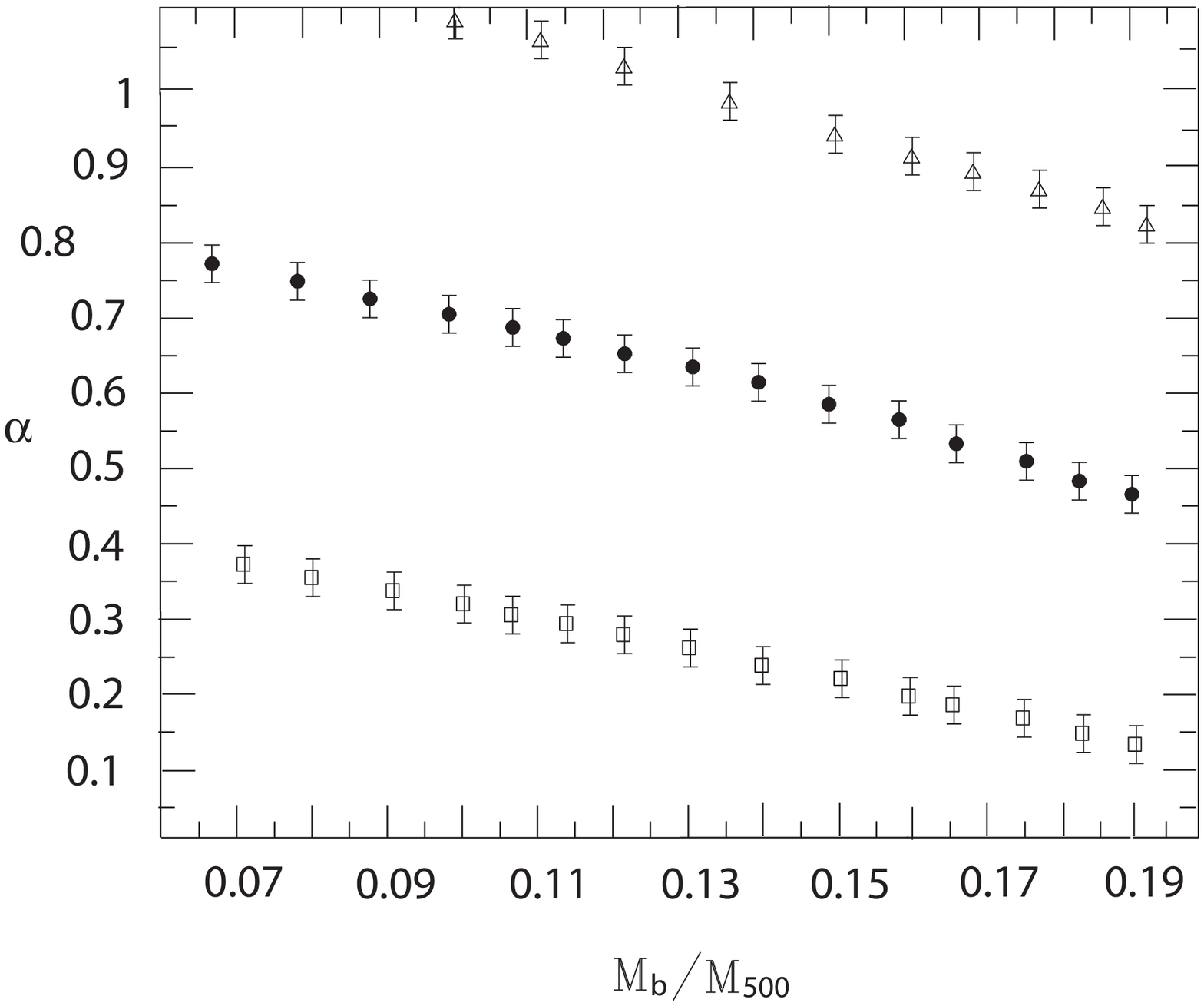}} (b) \goodgap \\
%\includegraphics[width=84mm]{dwarfsnmm.eps}
%\hspace{-0.15cm}
%\hspace{-0.5cm}
%\includegraphics[width=84mm]{dwarfs33.eps}
%\hspace{0.15cm} \includegraphics[width=68mm]{dwarfs4i.eps}
%\hspace{-0.15cm} 
%%%%%%%%%%%%%%%%%%%%%%%%%%%%%%%%%%%%%%%%%%%%%\subfigure{\includegraphics[width=6cm]{einas.ps}} \goodgap \\
%\hspace{0.15cm} \includegraphics[width=68mm]{dwarfs4ii.eps}
%\includegraphics[width=84mm]{dwarfsn_isol.eps}
\caption{Changes of the inner slope $\alpha$ in terms of the ratio of barionyc to total mass, $M_b/M_{500}$. Dots with error-bars refer to the case 
$f_d=f_{d_{\ast}}$, $j=j_{\ast}$. The triangles with 1 $\sigma$ error-bars refers to the case $f_d=f_{d_{\ast}}$, $j=j_{\ast}/2$, and the squares with error-bars to the case $f_d=f_{d_{\ast}}$, $j=j_{\ast} \times 2$.}
%Slope vs. massa barionica/massa totale. dati centrali ($F_b$,j), superiori $F_b$, j/2, inferiori $F_b$, $2 \times j$}
\end{figure*}
  
A comparison of Fig. 1a and Fig. 1b, shows that the $10^{14} M_{\odot}$ halo is less steep than the $10^{15} M_{\odot}$ one. 
%{\bf
The steepening of the cusp with mass is in agreement with previous studies (Cen \& Ostriker 2000; Ricotti 2003). 
%}
%Vedere Nipoti...MNRAS 2004, 355, 1119  , introduzione
Both Fig. 1a and Fig. 1b, show that haloes with higher angular momentum have smaller inner slopes. 
%In other terms: a) less massive haloes are less concentrated; b) the halo's inner slope is smaller for smaller mass. \\
%{\bf
The steepening of the profile with mass is due to the fact that higher density peaks (larger $\nu$\footnote{$\nu=\delta(0)/\sigma$, where $\delta$ is the mean fractional density excess inside a shell, and $\sigma$ is the mass variance filtered on a scale $R_f$ (see Appendix B of DP09).})
are usually progenitors of more massive haloes\footnote{As discussed by Peacock \& Heavens 1990; Del Popolo \& Gambera 1996 or Gao \& White (2007) (Fig.1), modelling the peaks as triaxial spheroids one obtains a peak mass $M=\frac{2^{3/2} (4 \pi/3) \rho_b R_{\ast}^3}{\gamma^3+(0.9/\nu)^{3/2}}$ for 
$0.5 \leq \gamma \leq 0.8 $ (where $\gamma$ and $R_{\ast}$ are spectral parameters given by Eq. (B6 and B7) in DP09). 
%are given in Eq. (\ref{eq:gammm}) and Eq. (\ref{eq:rrr}), 
The peak mass is an increasing function of $\nu$.
%showing an increase of mass with $\nu$. 
It is reasonable that lower $\nu$ peaks should have lower mass; peaks with $\nu \simeq 0$ will tend to sit in regions of larger scale underdensity (cancelling the small-scale overdensity), and hence the $\sim \rho_c R_{\ast}^3$ of material, which initially surrounds the peak, may not be accreted following central collapse. }
%(see Peacock \& Heavens 1990; Gao \& White 2007) 
which are characterized by a larger central density contrast. As a consequence, a generic shell inside this peak will fill more strongly the central potential and will expand less than if it was located in a smaller density peak. The angular momentum acquired during expansion will be reduced and haloes will be more concentrated. It is important to notice that the quoted trend of increased central concentration as a function of mass applies only to halos that started out as peaks in the density field smoothed with a fixed $R_f$ scale (see Appendix B of DP09). Our conclusions do not mean that, for example, clusters of galaxies will be more centrally concentrated than galaxies, since different smoothing scales would apply in the two cases.
%}
The angular momentum dependence of the slope arises from the fact that less massive objects are originated from peaks having a smaller height. They acquire less ordered, $h$, and random, $j$, angular momenta (Ryden \& Gunn 1987). The larger is the angular momentum, the more particles constituting the shell will remain closer to the maximum radius, producing a flatter density profile. Summarizing, particles having larger $h$ and $j$ tend to live far from the center, so not contributing to the central density. At the same time, a larger value of mass produces an increase of the central density contrast, and a steeper profile.

%In the following, we discuss 

%It is important to note as more widely discussed in the following, that the reason why larger angular momentum produces flatter profiles ??????? (da %togliere?)

The baryon content of the halo is another parameter playing a key role in shaping density profiles. In Fig. 1c, we studied how changes in the baryonic fraction change the profile of a $10^{14} M_{\odot}$ halo. Similar results, not plotted, are valid for $10^{15} M_{\odot}$ haloes. 

In Fig. 1c, the solid line is the same as Fig. 1a, while the short-dashed (dotted) line represents the case $f_d=f_{d_{\ast}}/1.5$ and $j=j_{\ast}$ ($f_d=f_{d_{\ast}}/1.5$ and $j=j_{\ast}/2$), where $f_{d_{\ast}}=F_{B_{\ast}}/f_b= 0.15/0.17=0.88$. The long-dashed (dot-dashed) line represents the case $f_d=f_{d_{\ast}} \times 1.5$ and $j=j_{\ast}$ ($f_d=f_{d_{\ast}} \times 1.5$ and $j=j_{\ast} \times 2$). 

The short-dashed line shows that a decrease of the baryonic fraction, from $f_{d_{\ast}}$ to $f_{d_{\ast}}/1.5$ produces a steepening of the density profile, to 
%In this case, 
$\alpha \simeq 0.7$.
If also $j$ is decreased, to $j_{\ast}/2$, a further steepening of the profile is produced ($\alpha \simeq 1.1$). In the reverse case, the long-dashed line, representing the case 
$f_d=f_{d_{\ast}} \times 1.5$ and $j=j_{\ast}$, the profile flattens ($\alpha \simeq 0.35$), and a further flattening of the profile is produced ($\alpha \simeq 0.1$ if also, $j$ is increased to $j=j_{\ast} \times 2$). 
The clear tendency to have flatter inner profiles for larger baryon content is connected to the fact that the presence of a larger quantity of baryons guarantees a larger transfer of energy and angular momentum from baryons to DM with the results that DM moves to larger orbits reducing the inner density.
It is important to notice, that an increase in angular momentum has a more important effect on the inner density profile slope than an increase in the baryon fraction (an increase of a factor of 2 in random angular momentum has a larger effect than an a similar increase in the baryon fraction).

In the case that the system does not contain baryons, like dissipationless N-body simulations, the density profile converge to the Einasto profile (as shown in Fig. 1d of DP09), characterized by a logarithmic slope that varies continuously with radius (Navarro et al. 2004, 2010).
%As shown in DP12 (Fig. 1d) (not replotted in this paper), in the case baryons are not present, the density profile is fundamentally an Einasto profile, %as the one predicted in dissipationless N-body simulations (e.g., Navarro et al. 2004, 2010). 

%{\bf
In Fig. 2, we plot how the negative logarithmic slope of the DM density at small radii ($10^{-2} r_{vir}$), $\alpha=\frac{d \log{\rho}}{d \log{r} }$,
varies with the baryonic content in haloes, $M_b/M_{500}$.
%}. 
The dots with 1 $\sigma$ error-bars represent the relation 
$\alpha$-$M_b/M_{500}$ in the case $f_d=f_{d_{\ast}}$ and $j=j_{\ast}$. The plot shows a decrease of $\alpha$ with increasing value of the baryonic fraction, $f_d$, as already noticed in Fig. 1: the final density profiles are shallower when $F_b$ (or similarly $f_d$) is higher. The triangles (squares) with 1 $\sigma$ error-bars, represent the same relation in the case $f_d=f_{d_{\ast}}$ and $j=j_{\ast}/2$ ($f_d=f_{d_{\ast}}$ and $j=j_{\ast} \times 2$). The plot shows that for a given $f_d$ one can obtain different slopes according to the value of random angular momentum of the system, and this suggests that the cluster mass and baryonic fraction are not the only fundamental parameters in shaping the density profile. A fundamental role is played by the orbital properties of the objects constituting the cluster. Galaxies moving on different orbits heat the DM at different rates. Galaxies having larger kinetic energy transfer, through dynamical friction, a larger amount of energy to DM. So, if baryon content has a certain importance in determining the final DM distribution, the orbital parameters of galaxies and their dynamics has a similar or even a larger importance.
%BISOGNA SPECIFICARE QUALE MODELLO USO!!!
%{\bf
In a recent paper, Laporte et al. (2012) studied how the evolution of galaxies influence clusters of galaxies formation. They followed galaxies evolution through collisionless mergers and found that 
%found that dissipationless mergers in simulations of galaxy clusters 
%as a result of mixing between stars and DM in dissipationless mergers, 
cusps with $0.8 <\alpha <1.3$ can be flattened to $0.3 <\alpha <0.9$ at the innermost resolved radii. This result is in qualitative agreement with the results of our paper. However, in their study, the density profiles depend not only on the orbital properties of galaxies, as in the present study, but also on  the internal structure of galaxies. 
%has also a certain importance in altering the dark matter profile. 
Less tightly bound galaxies (i.e., less compact) are more easily stripped with respect to more tightly bound ones, and as a result the more is the quantity of less compact galaxies in the cluster the more quantity of stellar mass is stripped and deposited on the BCG. The effect on the slope of the DM profiles is that clusters containing more compact galaxies have shallower inner profiles and vice versa. 
Even if the model is an improvement on that of El-Zant et al. (2004), it has several limitations as recognized by the authors, 
%As told by the authors, the result is a qualitative one, 
and certain aspects of matter distribution in their simulations are unrealistic. This is one of the reasons they did not even attempt to reproduce observations of clusters. Moreover, the flattening of the inner slope of the density profile of clusters is not only due to the effect of galaxy evolution. Dissipational processes have an important role in shaping galaxies at early times. Baryons are more important than DM in the inner parts of galaxy clusters and alter the DM profile in many ways, as described in the previous part of the present paper. 
%Finally, galaxy formation does not happen just through collisionless mergers.
%Moreover, galaxies do not form only through collisionless mergers, dissipational processes have an important role in shaping galaxies at early times, %baryons are more important than DM in the inner parts of galaxy clusters and alter the DM profile in many ways, as described in the previous part of the %present paper.

%}

%{\bf 

In Fig. 1, and Fig. 2, we discussed how angular momentum shapes the density profiles, and that haloes characterized by a larger angular momentum have flatter density profiles. This result could erroneously lead us to think that angular momentum is the primary effect driving the shaping of the density profiles, and the principal reason giving rise to the difference between density profiles found in simulations and those found in the present paper.
This is not the case. We have to recall that even if we are studying the effect of angular momentum change on the inner density slope, our system is composed of baryons and DM, and as a consequence our results are different from dissipationless N-body simulations. In our model, angular momentum is not working alone in shaping the density profiles, 
%As previously told, our system contains DM and baryons, and 
dynamical friction transfer angular momentum from baryons to DM. The larger is the angular momentum in the system, the larger is the quantity that can be transferred to DM with a consequent larger flattening of the density profile.  
I want also to add that the flattening of density profiles with increasing magnitude of angular momentum 
%when angular momentum is increased 
was also obtained in several studies even of systems not containing baryons (Avila-Reese et al. 1998, 2001; Subramanian et al. 2000; Nusser 2001; Hiotelis 2002; Le Delliou \& Henriksen 2003; Ascasibar, Yepes \& G\"ottleber 2004; Williams et al. 2004; Ascasibar, Hoffman \& Gottlober 2007). In peculiar, Ascasibar, Yepes \& G\"ottleber (2004), compared the analytical profiles of SIM which included non-radial motions with a set of high-resolution N-body simulations, showing that angular momentum is responsible for the shape of the density profile near the centre. Ascasibar, Hoffman \& Gottlober (2007) used N-body simulations to show that SIM gives density profile in good agreement with N-body simulations and that larger values of angular momentum produce a flattening of the profile. 

Angular momentum acts also on galaxies, and if it was the only effect taken into account in our model, it would delay their collapse, keeping them away from the center. In reality galaxies, as baryons, are also subject to dynamical friction which makes them loose their energy and sink to the center (Kashlinsky 1987).

The reason why our model gives different results from dissipationless N-body simulations was already discussed in DP09, and it is not connected to the way our model takes account of angular momentum.
%In summary, we may tell that
The issue is not if our model takes into account angular momentum more reliably than N-body simulations, but the difference among the system described in our model and that of dissipationless simulations.
% 
%%The reasons are several, as explained in DP09, but the main is connected to baryons presence and the interplay between baryons and DM (AC and exchange %%of angular momentum between baryons and DM through dynamical friction). 
%
%and here we summarize them.
In DP09 (beginning of page 2101), we described the two-fold effect of baryons presence, namely the steepening of the density profile due to AC and the flattening of the inner density profile due to the transfer of angular momentum from the
baryons to the DM. In collisionless N-body simulations, in which baryons are not present, this complicated interplay between different effects is not taken into account. Only recent SPH simulations (Romano-Diaz et al. 2008; Governato et al. 2010) took account of the baryonic component, finding the quoted cusp erasing.

%}

%{\bf
%We want to stress, that if the main reason why our results are different from those of dissipationless simulations is due to the different systems that %we are studying, a discussion of further reasons that could give rise to different results in dissipationless N-body simulations and analytical %treatments were discussed in DP09, (page 2098, and 2101).
%}

Finally, we recall that a check of the model against the Governato et al. (2010) SPH N-body simulations, has been performed in DP12 (Fig. 3).
%where we checked the model against Governato et al. (2010) SPH simulations.

\subsection{Comparison with observed clusters of galaxies}

%
%\begin{figure*}
%\centering
%\hspace{0.8cm}
%(a) 
%\subfigure{\includegraphics[width=7.1cm]{a611_color.eps}} (b)  
%\subfigure{\includegraphics[width=6.8cm]{a383_i.eps}} (c) 
%\subfigure{\includegraphics[width=6.6cm]{rxj1133_mio1111.eps}}   (d)
%\subfigure{\includegraphics[width=6.6cm]{morandi1.eps}}   
%\caption{Fig. 3a A611; Fig. 3b A383; Fig. 3c rxj1133; Fig. 3d macs....}
%\end{figure*}
%

\begin{figure*}
\centering
\hspace{0.8cm}
%\subfigure{\includegraphics[width=7.4cm]{ms2137_mio1.eps}} (b) \goodgap 
\subfigure{\includegraphics[width=15.1cm]{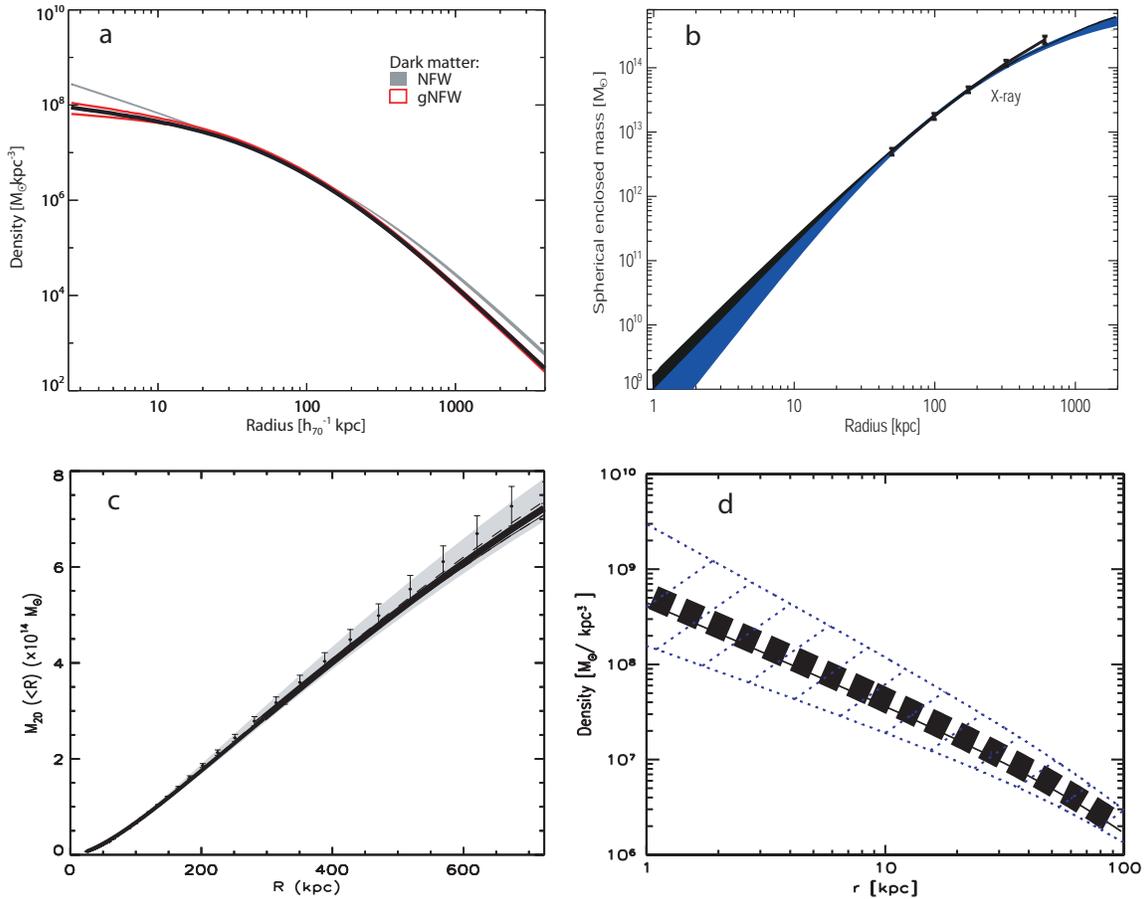}} 
\caption{Fig. 3a plots the density profile of A611. The grey solid line is the NFW profile, the red lines bracket the 68\% confidence region of the density profile obtained by N09, while the black band is the result of our model. Fig. 3b represents the mass profile of A383. The blue band is the observational result of N11, while
the black band the mass distribution obtained with our model. Fig. 3c represents the mass profile of MACS J1423.  
The azure band with errorbars is the observational result of Morandi, Pedersen \& Limousin (2010), while the black band is the result obtained with our model.  
Fig. 3d represents the density profile of RX J1133. The solid curve and the shaded region are the S04 observational result. The black dashed band is the result of our model. The confidence regions in Fig. 3a-c are 1 $\sigma$ while those in Fig. 3d are 2 $\sigma$.}
\end{figure*}

At this point, we compare the results of our model with density and mass profiles of observed clusters, ranging from a flat inner profile, namely that of 
%MS2137-23, 
A611, to intermediate ones, namely A383, to steep profiles, namely MACSJ1423.8+2404 and RXJ1347.5-1145.  

A611, A383, and RX J1133 are three of the clusters studied in a series of papers of Sand (S02, S04, and S08) and Newman (N09, N011).
%RX J1133 is one of the clusters studied in the series of papers of Sand (S02,S04, and S08; N09, N011),  
The quoted papers, followed a proposal of Miralda-Escude (1995), evidencing that if for a cluster we can add to the information on tangential and radial arc the information concerning the potential of the central galaxy (BCG), stronger constraints on the inner profiles of DM and luminous components can be obtained. 
S02 studied MS 2137-23, while S04 six clusters (MS2137-23, A383, A963, RX J1133, MACS 1206, and A1201), and S08 improved the results concerning MS2137-23 and A383.
N11 improved the constraints previously obtained in S08 concerning A383, and N09 studied A611.

%\subsubsection{MS 2137-23}

\subsubsection{A611}

A611 has been studied in several papers (OWLS team; Bonamente et al. 2004; Schmidt \& Allen 2007; Hurley-Walker et al. 2011 (AMI consortium)). N09, combined weak lensing from multicolor {\it Subaru} imaging, strong lensing ({\it Hubble Space Telescope}), and stellar velocity dispersion 
measures (Keck Telescope), sampling the dark matter profile from $\simeq 3$ kpc to $3.25$ Mpc. In order to accurately compare observational data and N-body simulations, one needs a) to separate DM from baryons, which, nevertheless they are a small part of the total mass of clusters, are usually dominating in the 
inner kpc scale; b) to probe mass on all scales. To this aim it is necessary to consider a wide dynamic range in radius. Concerning the mass distribution in the range 150 kpc- 3.25 Mpc, N09 found values of $r_s=320^{+240}_{-110}$ kpc, and 
%$c_{200}=5.1^{+1.7}_{-1.6}$ (scrivere cos'e' rs e c200)
$c=5.1^{+1.7}_{-1.6}$ (in agreement with Schmidt \& Allen 2007). The distribution of the stellar mass of BCG, measured through strong lensing and stellar velocity dispersions, is well fitted by a $R^{1/4}$ law in the 
3-20 inner kpc. The BCG is characterized by  $L_B =(5.6 \pm 0.8) \times 10^{11} M_{\odot}$, and $M_{\star, BCG}/L_B=2.7^{+0.7}_{-0.8}$ (N09). Combining weak lensing, strong lensing, and stellar kinematics data, then they concluded that $\alpha <0.3, (<0.56, <0.65)$ at 68\%, (95\%, 99\%) CL. Schmidt \& Allen (2007) found a value of the slope $\alpha =0.64^{+0.94}$.
An important result obtained by N09 is that using a subset of their data coming from just one technique (e.g., weak lensing, or strong lensing), the data are well fitted by a NFW model, but NFW model is unable to fit all data simultaneously. This imply the need to use several combined techniques and large radial scales. We compared the result of our model to N09 DM density profile in Fig. 3a. 
The cluster mass is estimated as $M_{vir}=6.18^{+3.82}_{-1.81} \times 10^{14} h^{-1} M_{\odot}$ (Schmidt \& Allen 2007), $M_{200}=5.6^{+4.7}_{-2.7} \times 10^{14} h^{-1} M_{\odot}$ (Romano et al. 2010). We recall, that, as previously reported, all masses in the paper were converted to $M_{500}$.
In order to determine the DM density profile, we used our model. 
%the Schmidt \& Allen (2007) mass and the baryonic fraction calculated by using Giodini et al. (2009), McGaugh et al. (2010), and from the Schmidt \& Allen (2007) data is $\simeq 0.15$ ?
The baryonic fraction obtained as described in the final part of Sect. 2, 
%using Giodini et al. (2009), McGaugh et al. (2010) are in agreement with the value obtained using Schmidt \& Allen (2007)\footnote{We recall that in this case, we calculated the mass of the cluster using the NFW parameters given in their Tab. 3 (relative to the total mass) and subtracting to the mass calculated in a similar way using the parameters given in their Tab. 4 (relative to the DM mass)}. The value of 
gives a value of $M_b/M_{500} \simeq 0.15$ ($f_d \simeq 0.88$). The value of the baryonic fraction is larger with respect to that of RX J1133, and MACS J1423 (as we shall see in the following), and this is one reason why one should expect a flatter inner profile with respect to the two quoted clusters, well fitted by a NFW model.  
In Fig. 3a, the grey line is the NFW fit, the red lines bracket the 68\% confidence region of the density profile obtained by N09, while the black band
%dashed line 
is the result of our model (68\% CL). 
%
%VALORI DI ALPHA?
%
%As told before the baryonic fraction of A383 is larger than the standard case $\simeq 0.1$ and this contribute to flatten the inner density profile with the standard case Fig. 1. 
Looking at Fig. 1b, for a cluster having a mass $\simeq 10^{15} M_{\odot}$ and $f_d=f_{d_{\ast}}$, and $j=j_{\ast}$, the inner slope is of the order of 0.7. Taking account the fact that A611 has a mass smaller than $10^{15} M_{\odot}$,
%of the smaller mass of A611, 
one would obtain a value of $\alpha \simeq 0.65$. In order to obtain the good fit plotted in Fig. 3a, we had to increase the magnitude of $j$ to $2 \times j_{\ast}$. As shown in Fig. 1, and also in Williams et al. (2004) (Fig. 6), increasing the value of $j$ produces a flattening of the profile. 
%Taking account of the smaller mass of A611 and of a $F_B=0.15$ which implies a smaller value of $\alpha$, of the order of 0.1), one obtains value of $\alpha \simeq ?5$, in agreement %with the the A611 profile obtained by Newman et al. (2009). 

\subsubsection{A383}

A383, is one of the most X-ray luminous clusters in the redshift slice $0.17<z <0.26$, at $z=0.189$. It is characterized by numerous strong lensing features and a dominant central galaxy. 
An analysis of the mass distribution in the core, built up through a lens model of the cluster, was carried out by Smith et al. (2001, 2005). The inner density profile, was evaluated by them to be $\alpha = 1.3 \pm 0.04$. 
%(probably the total mass). 
The cluster was one of the quoted six clusters studied in S04, combining results from stellar velocity dispersion data, and lensing. The cluster has both a radial and tangential arc. The dispersion velocity profile, and the density profile are plotted in Figs. 7, 8 of S04. The inner DM slope obtained by S04 was $\alpha=0.38^{+0.06}_{-0.05}(^{+0.12}_{-0.12})$ (68\% (95\%) CL). The results of S04, and consequently the constraints on A383 inner slope, were criticized by Meneghetti et al. (2007)\footnote{Notice that Meneghetti uses traxial clusters similar to those obtained in simulations, while those in S04 are more spherical.}, on the base that S04 used axially symmetric models in order to obtain a description of the six clusters mass model, while a pseudo-elliptical lens model (including ellipticity in the lensing potential) gives rise to steeper density profiles. The previous criticisms were tackled in S08, in the case of A383, finding again a flat inner DM slope ($\alpha= 0.45^{+0.2}_{-0.25}$).

A further improvement to the model was presented in N11, in which the authors, used the same technique used in N09.
%weak lensing from multicolor {\it Subaru} imaging, strong lensing ({\it Hubble Space Telescope}, and stellar velocity dispersion measures (Keck %Telescope), were employed, finding a dark matter profile from $\simeq 3$ kpc to $3.25$ Mpc. 
They also combined X-ray and lensing constraints, in order to measure the DM elongation along the line of sight. In this way, it was further possible to tackle an important systematic uncertainty in the mass profile determination (see Morandi et al. 2010). As a result, they further improved the S08 constraints, who has not taken into account triaxiality. They obtained a value of $\alpha<0.70$ (68\% confidence), $\alpha<1$ (95\% confidence), and a best fit (inferred from weak, strong lensing, kinematics, and X-ray data) of $\alpha=0.59^{+0.30}_{-0.35}$.\footnote{The authors are also studying other nine clusters with the same level of precision.} 
%PARTE DELLA PARTE SUCCESSIVA E' UGUALE A QUELLA DI A611
In order to compare the mass profile obtained by N11, with our model, we need mass and baryonic content of the cluster. Zitrin et al. (2011) and Schmidt et al. (2007) found values of $M_{vir} =6.26^{+0.26}_{-0.25} \times 10^{14} h^{-1} M_{\odot}$, and $M_{vir} =6.87^{+1.89}_{-1.85} \times 10^{14} h^{-1} M_{\odot}$, respectively. The mass of the BCG, within $\simeq 19$ kpc, is $1.14 \pm 0.03 \times 10^{12} M_{\odot}$ (Zitrin et al. 2011) in agreement with N11. The baryonic fraction was obtained as for A611.
%obtained using Giodini et al. (?), McGaugh et al. (2010) are in agreement with the value obtained using Schmidt et al. (2007)\footnote{We recall that in this case, we calculated the %mass of the cluster using the NFW parameters given in their Tab. 3 (relative to the total mass) and subtracting to the mass calculated in a similar way using the parameters given in %their Tab. 4 (relative to the DM mass)}. 
The value of $M_b/M_{500}$ is $\simeq 0.15$ ($f_d \simeq 0.88$). 
%The value of the baryonic fraction is larger with respect to RX J 1133, and this is one reason why one should expect a flatter inner prof
In Fig. 3b, we compare the mass distribution obtained with our model (black band), with that of N11 (blue band). 
%As told before the baryonic fraction of A383 is larger than the standard case $\simeq 0.1$ and this contribute to flatten the inner density profile with the standard case Fig. 1. %looking at Fig. 1b, for a cluster having a mass $\simeq 10^{15} M_{\odot}$ and $F_B=F_{B_{\ast}}$, $j=j_{\ast}$, the inner slope is of the order of 0.7. Taking account of the smaller %mass of A383 and of a $F_B=0.15$ which implies a smaller value of $\alpha$, of the order of 0.1), one obtains value of $\alpha \simeq 0.55$, in agreement with the A383 profile. 
The good fit to the mass distribution of A383 is obtained with the typical value $j_{\ast}$ of random angular momentum. This shows that even if 
A383 has a mass and baryonic fraction close to that of A611, the smaller value of $j$ give rise to the steeper profile observed ($\alpha \simeq 0.6$).
%but its smaller value of $\j$ implies a steeper profile (as observed) with $\alpha \simeq 0.6$. 

\subsubsection{MACSJ1423.8+2404}

The cluster of galaxies MACS J1423.8+2404 (in the following referred as MACS J1423), is a
triaxial \footnote{DM halo axial ratios are $1.53 \pm 0.15$, on the plane of the sky, and $1.44 \pm 0.07$, along the line of sight (Morandi, Pedersen \& Limousin 2010).} massive ($M=4.52^{+0.79}_{-0.64} \times 10^{14} M_{\odot}$, Schmidt \& Allen 2007)
strong cooling core source\footnote{i.e., the central cooling time is much smaller than the age of universe.} (Morandi et. al. 2007b) at $z=0.539$. It is very relaxed (Kartaltepe 2008) with very low central temperature ($\sim 2$ keV). Morandi, Pedersen \& Limousin (2010) performed a joint analysis (X-ray, strong, and weak lensing) to determine DM density profile and ICM parameters. The BCG mass is $5 \times 10^{11} M_{\odot}$.
The three-dimensional analysis of the cluster, lead (Morandi, Pedersen \& Limousin 2010) to measure an inner slope $\alpha=0.94 \pm 0.09$ smaller than the two-dimensional spherical modeling, giving a value of $\alpha=1.24 \pm 0.07$. Similarly to the previous clusters, we used our model to determine the mass profile of the cluster. The value of the baryonic fraction, inferred as in A383, and A611, gives $F_b \simeq 0.1$ ($f_d \simeq 0.59$), 
%(ma Schmidt da' 0.049…), 
and the mass profile, plotted in Fig. 3d (black band), in good agreement with Morandi, Pedersen \& Limousin (2010) (azure band with errorbars), was obtained assuming $j=j_{\ast} /1.5$. The effect of steepening of the profile with decreasing $j$ was pointed out by Williams et al. (2004) (their Fig. 6).
Reducing the angular momentum of dark matter particles, reduces random velocities which results in steeper central
density slopes.

\subsubsection{RX J1133}

RX J1133 accurately studied in S04,
%RX J1133 is one of the clusters studied in the series of papers of Sand (S02,S04, and S08; N09, N011),  
%following a proposal of Miralda-Escude (1995), evidencing that if for a cluster one can add to information on tangential and radial arc that
%concerning the potential of the central galaxy (BCG), then stronger constraints on the inner profiles of DM and luminous components can be fixed. 
%S02 studied MS 2137-23, while S04 six clusters (A611, A383, MS2137-23,.......), and S08 improved the results concerning MS2137-23 and A383.
%N11 improved the constraints previously obtained in S08 concerning A383, and N09 studied A611. 
is a cluster with total mass $\simeq 3.2 \times 10^{14} M_{\odot}$ (Cardone, Piedipalummbo, \& Tortora 2005; Lackner \& Ostriker 2010 (private communication)), and a BCG with mass $3 \times 10^{11} M_{\odot}$ 
%$5 \times 10^{11} M_{\odot}$? 
(Lackner \& Ostriker 2010 (private communication)).
%$2.5 \times 10^{12} M_{\odot}$ 
%(meglio $\simeq 10^{12} M_{\odot}$ . 
S04 identified a tangential and radial arc and measured the velocity dispersion profile of the BCG. S04 disentangled the DM and baryonic component, finding density profiles of DM, luminous matter, and total matter (see S04, Fig. 8). The inner slope of the DM profile was found consistent with that of a NFW profile ($\alpha=0.99^{+0.18}_{-0.14}$). 
In Fig. 3d, we plot the DM density profile of RX J1133. 
Fig. 3d represents the density profile of RX J1133. The solid curve and the shaded region is the S04 observational result. The black dashed band is the result of our model, 
%The dashed band is the S04 result, while the dashed line is the one that we obtained, 
using the previously given mass for the cluster, and the baryonic fraction calculated as for the previous three clusters.
%calculating the baryonic fraction according to Giodini et al. (2009), and McGaugh et al. (2010). 
The obtained value $F_b \simeq 0.1$ is smaller than that of A383 and A611, implying a steeper slope with respect to the two quoted clusters.
%$f_B \simeq f_{B_{\star}}$ ?, (0.68), on the entire length of the data, 
The model of this paper gives a very good fit to the DM density profile of RX J1133, obtained by S04, if $j=j_{\ast}/2$. 
The reduction of random angular momentum is justified as in the case of MACS J1423. 
Finally, we notice how the result shows that a NFW model is also a good fit to RX J1133 density profile.
%in agreement with NFW model.

The confidence regions in Fig. 3a-c are 1 $\sigma$ while those in Fig. 3d are 2 $\sigma$.

\section{Discussion}

\subsection{What are the origins of the differences?}

From the results of the previous sections, we arrive to two important conclusions, that we will discuss: a) not all clusters density profiles are fitted by the NFW model. Some of them have flat inner density profiles (e.g., A611), some have intermediate slopes, between pseudo-isothermal profiles and NFW profiles, (e.g., A383), and others are well fitted by NFW profiles (e.g., RX J1133); b) the $\Lambda$CDM model is unable to describe the density profile of some clusters. 
\footnote{Obviously issue a, and b are strictly connected. }

In connection with the first conclusion, a connected question is the following: what is the cause of the differences of density profiles in clusters? In the study of the role of baryons and random angular momentum developed in Sect. 3.1, we saw that the causes that give rise to different slopes in clusters are three, namely the mass (clusters with larger mass have steeper profiles), the baryon content (clusters richer in baryons have flatter profiles), and random angular momentum, $j$ (larger values of $j$ implies flatter profiles). We also saw that random angular momentum have a stronger role than baryons in shaping density profiles. A383 and A611 have a similar mass ($M_{vir}  \simeq 6 \times 10^{14} h^{-1} M_{\odot}$), and a similar baryonic fraction
($F_B \simeq 0.15$ ($f_d \simeq 0.88$)). RX J11333 and MACS J1423 have slightly smaller masses ($\simeq 3-4 \times 10^{14} h^{-1} M_{\odot}$), and smaller baryonic fraction 
($F_B \simeq 0.1$ ($f_d \simeq 0.59$)). The difference in baryonic fraction is one of the cause of the steeper profile of RX J11333 and MACS J1423 with respect with the other two clusters,
%have a steeper density profile with respect to A383 and A611, 
in agreement with observations. We should also not forget that, similarly to dwarf galaxies, in order two clusters have a similar density profile, it is not only important that the baryonic fraction is similar for the two objects, but the way baryons are distributed has also an important role.
%\footnote{For example, in the case of dwarf galaxies, Hoeft et al. (2006) showed in their Fig. 5 how the baryon fraction can largely change, going from %the centre to the outskirts of the dwarf galaxy.}

In particular, the larger or smaller concentration of gas and stars in the central $\simeq 10 $ kpc, in the BCG, is also important. As shown by Schmidt \& Allen (2007) (Fig. 4), the inner slope of clusters of galaxies (MS 2137.3-2353, in particular) decreases with increasing values of the central mass. In the inner $\simeq 10$ kpc, A611 has a $M_{BCG}/M_{vir} \simeq 0.004$, using the data of N09, and of Schmidt \& Allen (2007), concerning $M_{BCG}/M_{vir}$. A383 has $M_{BCG}/M_{vir} \simeq 0.002$ (within $\simeq 19$ kpc), using N11; Schmidt \& Allen (2007), and Zitrin et al. (2011), data. RX J1133 has  $M_{BCG}/M_{vir} \simeq 0.001$ (Cardone, Piedipalummbo, \& Tortora 2005; Lackner \& Ostriker 2010 (private communication)), and MACS J1423 has a similar value.  
The larger value of ratio of the inner stellar mass to total mass of A611 with respect to A383, and the larger ratio of A611, and A 383 with respect to RX J1133 also implies 
that the inner slope of A611 is flatter than that of A383, and flatter than those of MACS J1423, and RX J1133.
%$\alpha_{A383}< \alpha_{A383} < \alpha_{RXJ1133}$...
~\\

It is interesting to discuss more in detail the role of the total baryonic mass, $M_b$, and the central one. 
As shown in El-Zant et al. (2001, 2004), Romano-Diaz et al. 2008, DP09, Governato et al. (2010), baryons presence produces in general a flattening of the density profile. The baryon diffused component produces rounder halos and less triaxial halos than those seen in DM simulations (Gustafsonn et al. 2006; Debattista et al. 2008; Abadi et al. 2010), and this change of shape has also influence on dynamical mass estimate (N09). 
The final configuration of a cluster is fixed by the initial quantity of baryons present in the proto cluster and by collapse/formation process. 

%
%%When we observe a cluster like A383 or A611, we observe the final result of the formation process: a BCG in the center of the cluster is usually present, together with %%diffused baryons. At the beginning of the formation process, the system contained a certain quantity of baryons, and DM that after collapse gave rise to the BCG and all the 
%other cluster characteristics. In other words, the final configuration is fixed by the initial quantity of baryons present in the proto cluster. 
%%From the point of view
%%of SPH N-body simulations, or semi analytical models, the situation is the reverse: we fix the initial baryon content and this will give rise through the %%collapse/formation process to the final configuration. 
%%In this sense, in simulations (semianalytical calculations), it is more important the initial baryonic content of the protocluster.   ?????
%
\begin{figure*}
\centering
\hspace{0.8cm}
\subfigure{\includegraphics[width=15.4cm]{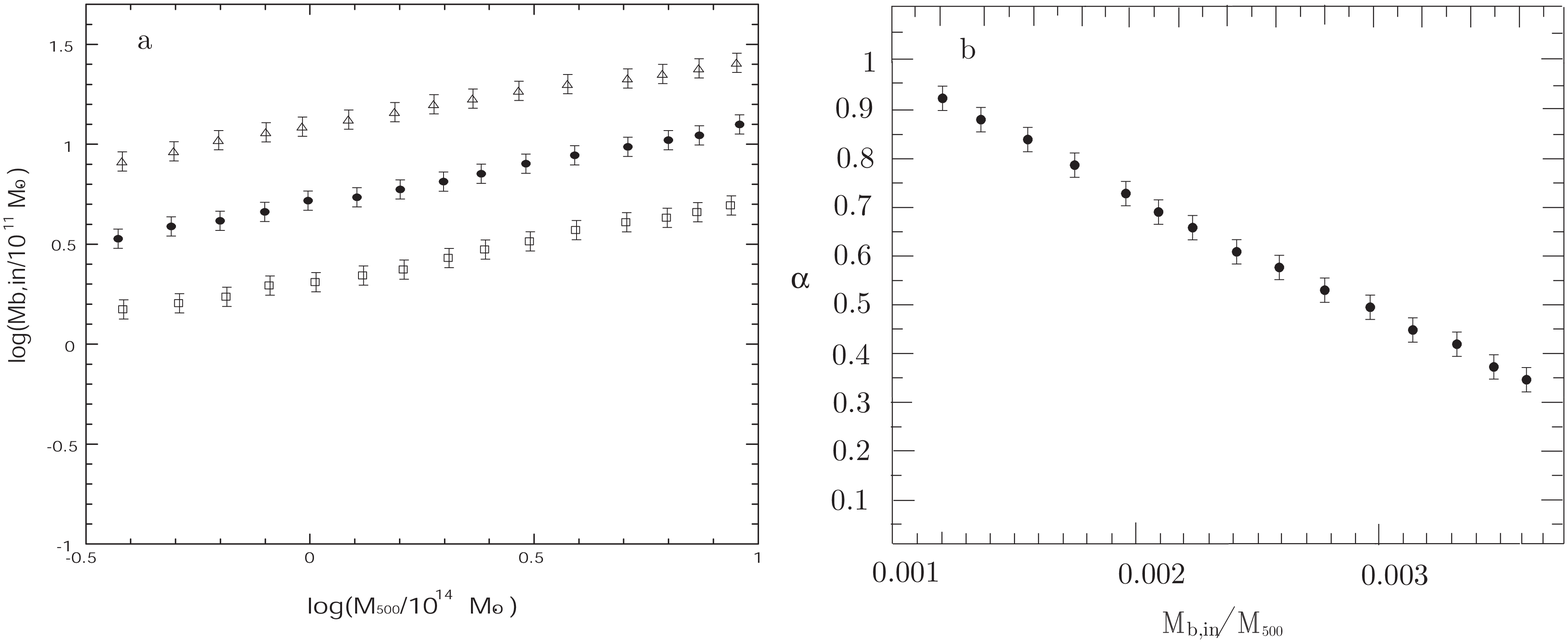}} \goodgap \\
\caption{Fig. 4a: baryonic mass in the inner 10 kpc in terms of the total mass. Dots with 1 $\sigma$ error-bars, triangles with errorbars, and squares with error-bars represent the quoted relation in the cases $f_d=f_{d_{\ast}}$ and $j=j_{\ast}$; $f_d=f_{d_{\ast}}$ and $j=j_{\ast} \times 2$; and $f_d=f_{d_{\ast}}$ and $j=j_{\ast}/2$, respectively. Fig. 4b: inner slope in terms of the ratio 
%$M_{BCG}/M_t$. 
$M_{b,in}/M_{500}$. 
%Error-bars are 1 $\sigma$ deviations.
Dots with error-bars represent the quoted relation in the case $f_d=f_{d_{\ast}}$ and $j=j_{\ast}$.} 
\end{figure*}

From this considerations, we should expect, in hierarchical formation models, that the final central baryonic content and the BCG mass is somehow correlated with baryonic and total cluster mass. This is in fact the case, as shown by Whiley et al. (2008), who using the models of de Lucia \& Blaizot (2007), found that $M_{BCG}\propto M_{cl}^{0.4}$ or $M_{cl}^{0.5}$ depending on the feedback model used. Also a correlation between BCG luminosity and cluster X-ray luminosity was found by several authors (Schombert 1988; Edge 1991; Edge \& Stewart 1991; Hudson \& Ebeling 1997). Whiley et al. (2008) measured the quoted correlation as $M_{BCG}\propto M_{cl}^{0.12\pm0.03}$ for K~band magnitudes inside a diameter of 37~kpc (radius of 13$h^{-1}$~kpc). Brough et al. (2008) found $L_{BCG}\propto M_{cl}^{0.11\pm0.10}$ at K~band
inside 12$h^{-1}$~kpc (several other results are given in Lin \& Mohr 2004; Popesso et al. 2007; Yang et al. 2008; Haarsma et al. 2010). 
%BIBLIOGRAFIA DI TUTTA QUESTA PARTE IN HAARSMA. 

In Fig. 4a, we calculated the baryonic mass
%\footnote{Constituted mainly by stars and gas} 
in the inner 10 kpc and plotted it against the total mass. The dots with 
1 $\sigma$ errorbars represent the quoted relation for $f_d=f_{d_{\ast}}$ and $j=j_{\ast}$. The triangles with errorbars  represent the quoted relation for $f_d=f_{d_{\ast}}$ and $j=j_{\ast} \times 2$. The squares with errorbars  represent the quoted relation for $f_d=f_{d_{\ast}}$ and $j=j_{\ast}/2$. 
%Scrivere che abbiamo calcolato la massa della BCG col modello e che i valori sono simili a quelli delle osservazioni e fare il 
%PLOT $M_{BCG}\propto M_{cl}^{0.4}$ al variare di $j$.
The plot shows a similar correlation as that found in Whiley et al. (2008) 
\begin{equation}
%\frac{M_{BCG}}{10^{11} M_{\odot}} \simeq 1.3 \times 10^6 \left( \frac{M_{cl}}{10^{14} M_{\odot}} \right)^{0.4}
\frac{M_{b,in}}{10^{11} M_{\odot}} \simeq 1.3 \times 10^6 \left( \frac{M_{500}}{10^{14} M_{\odot}} \right)^{0.4}
\end{equation}

Fig. 4b, plots the inner slope (defined as in Fig. 2) in terms of the ratio $M_{b,in}/M_{500}$. Dots with 1 $\sigma$ error-bars represent the quoted relation in the case $f_d=f_{d_{\ast}}$ and $j=j_{\ast}$. 
%This relation refers to the case $f_d=f_{d_{\ast}}$ and $j=j_{\ast}$, namely the dots with 1 $\sigma$ error-bars.

In Fig. 4b, it is also important to note that the fundamental parameter influencing the inner slope of the density profile is 
%$M_{BCG}/M_{vir}$ 
$M_{b,in}/M_{vir}$ and not just the baryonic mass 
%$M_{BCG}$.
$M_{b,in}$.
%alone is not telling all the story.  
In fact, if two clusters have the same values of 
%$M_{BCG}$, 
$M_{b,in}$,
but different $M_{vir}$, the one having smaller value of $M_{vir}$ would have a flatter profile, since the role of the central baryonic component is larger. This was also noticed by Schmidt \& Allen (2007) (Fig. 4), in the case of MS2137.3-2353, which is one of the least massive clusters in the sample that they used. They showed that increasing the values of the central stellar mass produces larger effects on the inner slope value, with respect to other cluster, because, as reported, the total mass of MS2137.3-2353 is smaller than the other sample components.

The other quantity of fundamental importance in shaping the density profiles is the random angular momentum. We have seen that a good fit to the mass/density profile of the studied clusters implies different values of $j$. Fig. 2 clearly shows that cluster mass and baryonic fraction are not the only fundamental parameters in building up the cluster structure. Random angular momentum, $j$, strongly influences cluster formation. During the collapse, particle follows orbits connected to the value of 
$j$ that they have, and particles endowed with larger kinetic energy will transfer more energy (through dynamical friction) to DM with the results that DM will expand reducing the inner density. So, as previously reported, galaxies orbital parameters, and dynamics have a fundamental role in density profile formation, even larger than the baryonic content.

\subsection{Problems for the $\Lambda$CDM}

In the previous subsection, as in the introduction, we pointed out that $\Lambda$CDM model is unable to explain the flat density profiles observed in dwarf galaxies, and 
to describe the density profile of some clusters, having too flat inner slopes with respect to the $\Lambda$CDM model predictions. However this problem does not necessarily imply a problem for CDM model (e.g. DP09; Governato et al. 2010). 
In the clusters that we studied, only RX J1133 is in agreement with $\Lambda$CDM model predictions, while the inner slope of the density profiles of A611 and A383 contradicts the N-body simulations results. Although more recent simulations (e.g. Navarro et al. 2010) found density profiles characterized by 
a continuous flattening of the inner slope, this does not solve the contradiction between observations and simulations, since the quoted clusters have smaller slopes than 
the minimum value of  the slope ($\alpha=0.8$ at 120 pc) found in high-resolution dissipationless N-body simulations (e.g., Stadel et al. 2009).
The quoted discrepancy is not restricted to the clusters studied in the present paper. On the observational side, as discussed in introduction, X-ray observations (Ettori et al. 2002; Arabadjis, Bautz \& Garmire 2002; Lewis, Buote \& Stocke 2003), lensing (Tyson et al. 1998; Smith et al. 2001; Dahle, Hannestad \& Sommer-Larsen 2003; S02; Gavazzi et al. 2003; Gavazzi 2005; S04; Brada{\v c} et al. 2008; Limousin et al. 2008), dynamics (Kelson et al. 2002; Biviano \& Salucci 2006), or studies combining several techniques (e.g. S02; S04; S08; N09, N11), lead to large scatter in the value of $\alpha$ from one cluster to another. If some scatter can be explained, as reported in introduction by different/limited dynamic range in radius in different studies, BCG role not taken into account, or as pointed out by Morandi, Pedersen \& Limousin (2010), to standard (simplified) spherical modeling of clusters, as well as the degeneracy of $\alpha$ with parameters like $c$ and $r_s$, it is difficult to explain the large slope difference between some clusters (e.g., A611). 
%and those obtained in N-body simulations. 
If the scatter in the inner slope of density profile is not merely caused by limits in the techniques used, some other reason causes it.
%we should find a reason for these differences. 
We could conclude that either some assumptions of $\Lambda$CDM model are incorrect, or given the several and noteworthy evidences supporting $\Lambda$CDM on large scales, another possibility, moreover studied in DP09, is that baryonic physics is a fundamental issue in clusters formation. Probably, as noticed in DP09, the quoted discrepancy is merely due to the fact that we are comparing two totally different systems: the one generated by dissipationless simulations, not including baryons, with real structures whose 
physics is not just the dissipationless physics typical of DM.
%are trying to force dissipationless simulations to predict the same behavior for the density profile of a system whose physics is not just the dissipationless physics %typical of DM. }

To start with, we would recall that dissipationless simulations does not include baryons, which are usually dominant in the inner part of clusters\footnote{For example, within 10 kpc, the total density distribution is dominated by the BCG in A383, RX J1133, A1201, A963, MACS 1206, MS2137-23 (S04).}, and its presence strongly influence the DM distribution.
%it is not at all clear how their presence influence the DM distribution. 
If stars form earlier than DM, baryons will compress the DM (adiabatic contraction), giving rise to steeper profiles (Blumenthal. et al. 1986; Gnedin et al. 2004; Gustafsonn et al. 2006). However, adiabatic contraction can be counteracted by heating of DM due to dynamical friction with cluster galaxies (El-Zant et al. 2001, 2004; 
Nipoti et al. 2003; Romano-Diaz et al. 2008). 
Several studies have tried to rescue the $\Lambda$CDM paradigm without drastic changes to the $\Lambda$CDM physics, and studying baryon physics and stellar processes in the inner parts of galaxies and clusters. Interactions of the DM with a stellar bar (Weinberg \& Katz 2002; McMillan \& Dehnen 2005), decay of binary black hole orbits after galaxies merge (Milosavljevi\'c \& Merritt 2001), baryon energy feedback from active galactic nucleus (Peirani et al. 2008), the already quoted dynamical friction of stellar/DM clumps against the background DM halo (El-Zant et al. 2001, 2004; Romano-Diaz et al. 2008, 2009), random bulk motions of gas in primordial galaxies, driven by supernova explosions (Mashchenko et al. 2006), removal of low-angular momentum gas (Governato et al. 2010), are some of the solutions proposed. 
Moreover, Zappacosta et al. (2006) concluded, through X-ray observations of A2589, that processes in clusters of galaxies counteract adiabatic contraction.
Infalling DM subhaloes, according to their mass, could steepen or flatten the DM cusp (Ma \& Boylan-Kolchin 2004), and Nipoti et al. (2004) got similar results according to the baryon fraction of infalling galaxies. 
%Halo triaxiality has also effects on observations (VEDI PAGINA 13 di s07)..............

Another important caveat is that density profiles are not universal, and that in nature may exist a distribution of inner slopes broader than that
predicted by dissipationless N-body simulations. Every cluster has its own formation/merger history leading to different cluster characteristics (see Navarro et al. 2010, and S04). Several studies argue against universality of DM profiles (Jing \& Suto 2000; Subramanian 2000; Ricotti 2003, 2004, 2007; Cen et al. 2004; Simon et al. 2005; Merrit et al. 2005; Graham et al. 2006; Schmidt et al. 2008; Del Popolo 2009; Ma et al. 2009; Host \& Hansen 2010; Del Popolo 2012). N-body simulations as Jing \& Suto (2000), Fukushige et al. (2003), Ricotti (2003), Ricotti \& Wilkinson (2004) found inner slope variations from run-to-run, or with mass. Even Navarro et al. (2004, 2010) found dependence of the inner slope with mass, interpreted as a reflection of the trend between the concentration of a halo and its mass. Gao et al. (2008) in two very large cosmological simulations, found that density profiles deviate slightly but systematically from the NFW form and are better approximated by an Einasto profile. Moreover, the shape parameter of the quoted profile changes with mass and redshift. Merrit et al. (2005, 2006) interpret the variation in profile shape with halo mass, as an indication of the fact that $\Lambda$CDM halos have not a really universal profile, as already claimed by all the authors previously quoted. 
%Jing \& Suto 2000; Subramanian et al. 2000; Ricotti 2003, Ricotti \& Wilkinson 2004; Cen et al. 2004; Merrit et al. 2005; Ricotti et al. 2007; Host \& Hansen 2009. 
By using DP09, we studied, in Del Popolo (2011), the pseudo phase-space density, arguing against universality of density profiles constituted by dark matter and baryons. From the observational point of view, Simon et al. (2005), S02, S04, S07 studies, argue against universality of density profiles. 

If density profiles are not universal, is of fundamental importance to collect further high quality data, like those in A611 or A383, for larger samples of clusters (see N09, and N11) to calculate not only the mean of the distribution but also their moments. At the same time, SPH simulations, similar to that of Governato et al. (2010) for dwarfs, 
could be run to study clusters formation and evolution, then comparing the results with the measured distribution of inner slopes. 
%Given the centrality of this problem
%and the success of CDMat other scales, this seems aworthy goal
%to pursue, although its solution will likely require advances in
%computing power, algorithms, and understanding of the relevant
%physics.

\section{Conclusions}

In the present paper we have studied how changes in random angular momentum and baryonic fraction affect density profiles of clusters with masses in the range $10^{14}-10^{15} M_{\odot}$. The paper extends the study of DP12, on dwarf galaxies, to cluster of galaxies. 
%the study of DP12 concerning dwarf galaxies. 
A reference density profile was calculated using DP09, and then we
studied how density profile changes when changing baryonic fraction and angular momentum. Similarly to the case of dwarfs, the inner density profile steepens with increasing value of the halo mass, and with decreasing values of angular momentum and baryonic fraction (see Figs. 1 and 2). 
This is due to the fact that when more baryons are present the energy and angular momentum transfer from baryons to DM is larger, and DM moves on larger orbits reducing the inner density. Haloes constituted only of DM have Einasto's density profiles.  
The previous calculation was applied to four clusters (A611, A383, MACS J1423, and RX J1133). A611 has a flat inner profile, as found by N09, and a baryonic fraction 
$F_b \simeq 0.15$ ($f_d \simeq 0.88$). Its density profile is re-obtained with a random angular momentum $2 \times j_ {\ast}$. A383 has a flat inner profile, but steeper than that of A611. It has a similar baryonic fraction and we re-obtain its mass distribution with the typical value $j_{\ast}$.
MACS J1423 and RX J1133 have steeper profiles, with RX J1133 well fitted by a NFW profile. They are characterized by a smaller baryonic fraction of A611 and A383, 
$F_b \simeq 0.1$ ($f_d \simeq 0.59$) and their profiles are re-obtained with $j=j_{\ast} /1.5$, and $j=j_{\ast} /2$, respectively. 
Since the baryonic content in the inner kpcs of clusters can influence their inner slope, as described by Schmidt \& Allen (2007),
we studied how the baryonic mass in the inner $\simeq 10$ kpc relates with the total mass of the cluster. We found a similar correlation to that found by Whiley et al. (2008), namely an increase of central baryonic mass with total cluster mass. Clusters having larger central baryonic mass have flatter profiles. 
We also found that a fundamental role is played by the orbital properties of the objects constituting the cluster. 
So, if baryon content has a certain importance in determining the final DM distribution, the orbital parameters of galaxies, constituting the cluster, and their dynamics have a similar or even a larger importance.
In summary, the density profile of clusters is strongly influenced by baryons, random angular momentum, and from the orbital parameters of galaxies, constituting the cluster, and their dynamics.

%Differently from DP12, we did not study the effects of environment on clusters inner slope. 
%the merging history.
%and their shape (real shape and projection effects). 

Differently from DP12, in this paper we did not study the eventual correlation of density profiles with environment.
%An important issue that we did not study in this paper is the eventual correlation of density profiles with environment, has done in DP12 for dwarf %galaxies. 
In that paper, we found that environmental effects influence density profiles of dwarf galaxies. 
It would be interesting to deal with this issue in a future paper, even if this study is more complicated than in the case of dwarf galaxies. 
This study has a fundamental complication: in order to study an eventual environment- density profile correlation, we need a large sample of clusters from which to extract
%clusters that are not strongly interacting with neighbours, and to compare with dwarf galaxies in voids or to wait for more sophisticated
%techniques
isolated and unisolated clusters. From an observational point of view, there have not been many studies on the effects of the environment on the slopes
of the dark matter haloes (but see DP12).
%, simply because there are not yet enough
%large samples, which are necessary for this.
Moreover, the concept of isolated or un-isolated is more or less well defined for dwarfs (e.g., Karachentsev et al. 2004, Karachentsev \&  Kashibadze 2006), and some samples, even if not large, are already present in literature (WHISP; Swaters et al. 2002) and in the next few years some others will be added (the LITTLE THINGS 
%(PI: D. Hunter) 
and VLA-ANGST), 
%(PI: J. Ott)), 
for the case of clusters the situation is more complicated. 
To our knowledge, in literature Plionis et al. (2009) identified sub-clusters within 2500 km/s of the main cluster, deliberately excluding those cluster with distorted X-ray morphology and other indications of strong interactions. Hence their sample is conservative and we would expect them to be non-interacting in a significant way with neighboring ones (even if we cannot rule out that they are weakly interacting with a structure that isn't delineated due to sub-sampling of redshifts).
From that, Pimbblet (2010) constructed another restricted, and more isolated sample.
%sample, also excluding those clusters with a structure "contamination flag" from Miller et al (2005).  
 
Clearly, much theoretical and observational work needs to be done in order to obtain insights into the role of the environment on the density profiles of 
cluster of galaxies.
%dwarf galaxies.

%\end{document}

\section*{Acknowledgments}
%\acknowledgements

We would like to thank Tommaso Treu, Robert Schmidt, David Sand, Alister W. Graham, Claire Lackner, Andrea Morandi, Crescenzo Tortora, Igor Karachentsev, Valentina Karachentseva, Stefano Ettori, for stimulating discussions on the topics related to the subject of this paper. Finally we thank 
the referee, A. B. Newman for providing constructive comments and help in improving the contents of this paper. 

\bibliographystyle{mn2e}
\bibliography{paper}

\begin{thebibliography}{}
\bibitem{} Aarseth S.J., Binney J., 1978, MNRAS 185, 227
\bibitem{} {Abadi}, M.~G., {Navarro}, J.~F., {Fardal}, M., {Babul}, A.,  \& {Steinmetz}, M. 20010, MNRAS 407, 435–-446
%arXiv:0902.2477
\bibitem{} Arabadjis, J.~S., Bautz, M.~W., Arabadjis, G. 2004, ApJ 617, 303
\bibitem{} Arabadjis, J.~S., Bautz, M.~W., Garmire, G.~P. 2002, ApJ, 572, 66
\bibitem{} Ascasibar Y., Yepes G., Gottl¨ober S., 2004, MNRAS, 352, 1109
\bibitem{} Ascasibar Y., Hoffman Y., Gottl¨ober S., 2007, MNRAS, 376, 393
\bibitem{} Avila Reese V., Firmani C., Hernandez X., 1998, ApJ, 505, 37
\bibitem{} Avila-Reese, V., Firmani, C., Klypin, A., \& Kravtsov, A. 1999, MNRAS, 310, 527
\bibitem{} Avila-Reese, V., Colin, P., Valenzuela, O., D'Onghia, E., \& Firmani, C., 2001, ApJ, 559, 516
\bibitem{} Bardeen J.M., Bond J.R., Kaiser N., Szalay A.S., 1986, ApJ 304, 15
\bibitem{} Barnes, J., \& Efstathiou, G. 1987, ApJ, 319, 575
\bibitem{} Barrow, J.D., Silk, J., 1981, ApJ 250, 432
\bibitem{} Bartelmann M., Meneghetti M., 2004, A\&A, 418, 413 
\bibitem{} Bett, P., Eke, V., Frenk, C. S., Jenkins, A., Helly, J., Navarro, J., 2007, MNRAS 376, 215
\bibitem{} Binney J., Silk J., 1979, MNRAS 188, 273
\bibitem{} Blais-Ouellette S., Amram P., Carignan C., Swaters R., 2004, A\&A, 420, 147
\bibitem{} Bonamente, et al. 2004, ApJ 614, 56-63
\bibitem{} Brada{\v c}, M., et~al. 2005, \aap, 437, 49
\bibitem{} Brada{\v c}, M., et~al. 2008, ApJ 681, 187
\bibitem{} Brough, S., Couch, W.~J., Collins, C.~A., Jarrett, T., Burke, D.~J., \& Mann, R.~G. 2008, MNRAS, 385, L103
\bibitem{} Cardone, V. F., Piedipalumbo, E., Tortora, C., 2005A\&A 429, 49
\bibitem{} Cardone, V. F., \& Sereno, M. 2005, A\&A, 438, 545 2003, ApJ, 593, 26
\bibitem{} Cen, R. Y., Dong, F., Bode, P., \& Ostriker, J. P. 2004, arXiv:astro-ph/0403352
\bibitem{} Dahle H., Hannestad S., Sommer-Larsen J., 2003, ApJ, 588, L73 
\bibitem{} {Debattista}, V.~P., {Moore}, B., {Quinn}, T., {Kazantzidis}, S., {Maas}, R., {Mayer}, L., {Read}, J.,  \& {Stadel}, J. 2008, \apj, 681, 1076
\bibitem{} de Blok W. J. G., Walter F., Brinks E., Trachternach C., Oh S-H., Kennicutt R. C., 2008, AJ, 136, 2648 
\bibitem{} de Blok, W. J. G., \& Bosma, A. 2002, A\&A, 385, 816
\bibitem{} de Blok, W. J. G., Bosma, A., \& McGaugh, S. 2003, MNRAS, 340, 657
\bibitem{} {De~Lucia}, G. \& Blaizot, J. 2007, MNRAS, 375, 2
\bibitem{} de Naray R. K., McGaugh S. S., de Blok W. J. G., 2008, ApJ, 676, 920
\bibitem{} de Naray R. K., McGaugh S. S., Mihos J. C., 2009, ApJ, 692, 1321
\bibitem{} Del Popolo, A., \& Gambera, M. 1996, A\&A, 308, 373
\bibitem{} Del Popolo, A., 2009, ApJ 698, 2093
\bibitem{} Del Popolo, A., 2010, MNRAS 408, 1808
\bibitem{} Del Popolo, A., 2011, JCAP 07, 014
\bibitem{} Del Popolo, A., 2012, MNRAS 419, 971
\bibitem{} Dressler A., 1978, ApJ 243, 2
\bibitem{} Edge, A.~C. \& Stewart, G.~C. 1991, MNRAS, 252, 428
\bibitem{} El-Zant, A. A., Hoffman, Y., Primack, J., Combes, F.,\& Shlosman, I. 2004, ApJ, 607, L75
\bibitem{} El-Zant, A. A., Shlosman, I., \& Hoffman, Y. 2001, ApJ, 560, 636
\bibitem{} Ettori, S., Fabian, A.~C., Allen, S.W., \& Johnstone, R.M., 2002, MNRAS, 331, 635
\bibitem{} Flores, R. A., \& Primack, J. R. 1994, ApJ, 427, L1
\bibitem{} Fukushige, T., Kawai, A., Makino, J., 2004 ApJ 606, 625-634
\bibitem{} Gao, L., \& White, S. D. M. 2007, MNRAS, 377, 5
\bibitem{} Gao L., Navarro J. F., Cole S., Frenk C. S., White S. D. M., Springel V., Jenkins A., Neto A. F., 2008, MNRAS, 387, 536
\bibitem{} Gavazzi R., 2005, A\&A, 443, 793G
\bibitem{} Gavazzi R., Fort B., Mellier Y., Pell{\' o} R., Dantel-Fort M., 2003, A\&A, 403, 11 
\bibitem{} Gentile, G., Salucci, P., Klein, U., Vergani, D., \& Kalberla, P. 2004, MNRAS, 351, 903
\bibitem{} Gnedin, O. Y., Kravtsov, A. V., Klypin, A. A., \& Nagai, D. 2004, ApJ, 616, 16
\bibitem{} Gottl\"ober, S. \& Yepes, G., 2007, ApJ 664:117-122
\bibitem{} Governato, F., et al., 2010, Nature 463, 203
%C. Brook2, L. Mayer3, A. Brooks4, G. Rhee5, J. Wadsley6, P. Jonsson7, B. Willman9, G. Stinson6, T. Quinn1 & P. Madau8
\bibitem{} Graham, A.W., Merritt, D., Moore, B.,  Diemand J., and Terzic, B., 2006, AJ 132, 2701
\bibitem{} Graham, A.W., Merritt, D., Moore, B., Diemand, J., and Terzic, B., 2006, AJ 132, 2685
\bibitem{} Gregory S.A., Tiffit W.G., 1976, ApJ 205, 716 
\bibitem{} Gustafsson, M., Fairbairn, M., \& Sommer-Larsen, J. 2006, Phys. Rev. D, 74, 123--522
\bibitem{} Haarsma, D.B., et al. 2010, ApJ 713, 1037–-1047
\bibitem{} Hayashi E. et al., 2004, MNRAS, 355, 794 
\bibitem{} Hayashi, E., et al. 2004, MNRAS, 355, 794
\bibitem{} Hiotelis N., 2002, A\&A, 383, 84
\bibitem{} Host, O., and Hansen, S.H., 2011, ApJ, 736, 52
%arXiv:0907.1097 [SPIRES].
\bibitem{} Hu, W., Kravtsov, A.V., 2003, ApJ 584, 702
\bibitem{} Hudson, M.~J. \& Ebeling, H. 1997, ApJ, 479, 621
\bibitem{} Hurley-Walker et al., 2011, arXiv: 1101.5912
\bibitem{} Icke V., 1973, A\&A 27, 1
\bibitem{} Jing, Y. P., \& Suto, Y. 2000, ApJ, 529, L69
\bibitem{} Kashlinsky, A., 1987, ApJ 312, 497
\bibitem{} Karachentsev I. D., Karachentseva V. E., Huchtmeier W. K., Makarov D. I., 2004, AJ, 127, 203
\bibitem{} Karachentsev I. D., Kashibadze O. G., 2006, Astrophysics, 49, 3
\bibitem{} Keeton, C. R. 2001, ApJ, 561, 46
\bibitem{} Kelson et al. 2002, ApJ 576, 720 
\bibitem{} Klypin, A., Kravtsov, A. V., Bullock, J. S., \& Primack, J. R. 2001, ApJ, 554, 903
\bibitem{} Klypin, A.,  Zhao, H-S., and Somerville R.S., 2002, ApJ 573, 597
\bibitem{} Kneib J.P., et al., 2003, ApJ, 598, 804 
\bibitem{} Komatsu, E., Smith, K.M., Dunkley, J., Bennett, C.L., Gold, B. et al., 2011, ApJS, 192, 18
\bibitem{} Kowalski, M., Rubin, D., Aldering, G., Agostinho, R.J., Amadon, A. et al.,2008, ApJ, 686, 749
\bibitem{} Kravtsov, A. V., Klypin, A. A., Bullock, J. S., \& Primack, J. R. 1998, ApJ, 502, 48
%\bibitem{} Kuzio de naray 2008 2009..........
\bibitem{} Lackner, C. N., Ostriker, J. P., 2010, ApJ 712, 88-100
\bibitem{} Laporte, C. F. P., White, S. D. M, Naab, T., Ruszkowski, M., Volker Springel, V., 2012, arXiv:1202.2357
\bibitem{} Le Delliou M., Henriksen R. N., 2003, A\&A, 408, 27
\bibitem{} Lewis, A.~D., Buote, D.~A., Stocke, J.~T., 2003, ApJ 586, 135L
\bibitem{} Limousin, M., et~al. 2008, \aap, 489, 23
\bibitem{} Lin C.C., Mestel L., Shu F.H., 1965, ApJ 142, 1431
\bibitem{} Lin, Y.T. \& Mohr, J.~J. 2004, ApJ, 617, 879
\bibitem{} Loeb, A., Peebles, P. J. E., 2003, ApJ 589, 29
\bibitem{} Lukic, Z., 2009, ApJ, 692, 217–228
\bibitem{} Ma, C.P., Chang, P.  and Zhang, J.,  arXiv:0907.3144 
%[SPIRES].
\bibitem{} Mahdavi, A., {Hoekstra}, H., {Babul}, A., {Sievers}, J., {Myers}, S.~T.,  \& {Henry}, J.~P. 2007, \apj, 664, 162
%\bibitem{} Massimiliano Bonamente,1, 2 Marshall K. Joy,2 John E. Carlstrom,3, 4 Erik D. Reese,5, 6 and Samuel J. LaRoque3, 4, 2004, ApJ 614, 56-63
\bibitem{} Mashchenko S., Couchman H. M. P., Wadsley J., 2006, Nat, 442, 539
\bibitem{} McMillan, P. J., \& Dehnen, W. 2005, MNRAS, 363, 1205
\bibitem{} Mellier Y., 1999, ARA\&A, 37, 127 
\bibitem{} Merritt, D., Navarro, J.F., Ludlow, A., and Jenkins, A., 2005, ApJ 624, L85 
%\bibitem{} Meneghetti, Massimo; Bartelmann, Matthias; Jenkins, Adrian; Frenk, Carlos,  2007, MNRAS 381, 171M
\bibitem{} Meneghetti, M.,  Bartelmann, M., Jenkins, A., Frenk, C., 2007, Volume 381, Issue 1, pages 171-186
\bibitem{} Miller C. J., et al., 2005, AJ, 130, 968
\bibitem{} Milosavljevi´c, M., \& Merritt, D. 2001, ApJ, 563, 34
\bibitem{} Miralda-Escud\'e, J, 1995, ApJ 438, 514
\bibitem{} Mo, H. J., Mao, S., \& White, S. D. M. 1998, MNRAS, 295, 319
\bibitem{} Moore, B. 1994, Nature, 370, 629
\bibitem{} Moore, B., Governato, F., Quinn, T., Stadel, J., \& Lake, G. 1998, ApJ, 499, L5
\bibitem{} Morandi, A., Pedersen, K., \& limousin, M., 2010, ApJ 713, 491
\bibitem{} Navarro et al. 2010, MNRAS 402, 21–34
\bibitem{} Navarro, J. F., et al. 2004, MNRAS, 349, 1039
\bibitem{} Navarro, J. F., Frenk, C. S., \& White, S. D. M. 1996, ApJ, 462, 563
\bibitem{} Navarro, J. F., Frenk, C. S., \& White, S. D. M. 1997, ApJ, 490, 493 
\bibitem{} Newman, Andrew B.; Treu, Tommaso; Ellis, Richard S.; Sand, David J., 2011, ApJ 728, 39
\bibitem{} Newman, Andrew B.; Treu, Tommaso; Ellis, Richard S.; Sand, David J.; Richard, Johan; Marshall, Philip J.; Capak, Peter; Miyazaki, Satoshi, 2009, ApJ 706, 1078
\bibitem{} Nusser A., 2001, MNRAS, 325, 1397
\bibitem{} Oh, K. S. 1990, PhD thesis, Univ. California-Santa Cruz
\bibitem{} Oh S-H., Brook C., Governato F., Brinks E., Mayer L., de Blok W. J. G., Brooks A., Walter F., 2010, AJ 142, 24 
\bibitem{} Padmanabhan, T., 1993, Structure formation in the universe (Cambridge: Cambridge University Press)
\bibitem{} Peacock J.A., Heavens A.F., 1985, MNRAS 217, 805
\bibitem{} Peacock, J. A., \& Heavens, A. F. 1990, MNRAS, 243, 133
\bibitem{} Peirani, S., Kay, S., \& Silk, J. 2008, A\&A, 479, 123
\bibitem{} Percival, W.J., Reid, B.A., Eisenstein, D.J., Bahcall, N.A., Budavari, T. et al., 2010, MNRAS, 401, 2148
\bibitem{} Pimbblet, K. A., MNRAS 411, 2637–2643
\bibitem{} Plionis, M., Tovmassian, H. M., Andernach, H., 2009, MNRAS 395, 2
\bibitem{} Popesso, P., Biviano, A., B{\"o}hringer, H., \& Romaniello, M. 2007, A\&A, 464, 451
\bibitem{} Power, C., Navarro, J. F., Jenkins, A., Frenk, C. S., White, S. D. M., Springel, V., Stadel, J., \& Quinn, T. 2003, MNRAS, 338, 14
\bibitem{} Ricotti, M., and Wilkinson, M.I., 2004, MNRAS 353, 867 
\bibitem{} Ricotti, M., 2004, MNRAS 2003, 344, 1237
\bibitem{} Ricotti, M., Pontzen, A.,  and Viel, M.,  2007, ApJ 663, 53
\bibitem{} Ryden, B. S., 1988, ApJ 329, 589
\bibitem{} Rix, H.-W., de Zeeuw, P. T., Cretton, N., van der Marel, R. P., \& Carollo, C. M. 1997, ApJ, 488, 702 
\bibitem{} Romano A., et al., 2010, A\&A, 514, 88
\bibitem{} Romano-Diaz, E., Shlosman, I., Heller, C.,\& Hoffman, Y. 2009, ApJ, 702, 1250
\bibitem{} Romano-Diaz, E., Shlosman, I., Hoffman, Y.,\& Heller, C. 2008, ApJ, 685, L105
\bibitem{} Rood H.H., Page T.L., Kintner E.C., King I.R., 1972, ApJ 175, 627
\bibitem{} Ryden, B. S., \& Gunn, J. E. 1987, ApJ, 318, 15 
\bibitem{} Salvador-Sol\'e E., Solanes J.M., 1993, ApJ 417, 427
\bibitem{} Sand D.~J., Treu T., Ellis R.~S., 2002, ApJ, 574, L129 
\bibitem{} Sand D.~J., Treu T., Smith G.~P., Ellis R.~S., 2004, ApJ, 604, 88 
\bibitem{} Sand, D. J., Treu, T., Ellis, R. S., Smith, G. P., Kneib, J.-P, 2008, ApJ 674, 711
\bibitem{} Schmidt, K.B., Hansen, S.H., and Maccio', A.V., 2008, ApJ 689, L33 
\bibitem{} Schmidt, R.~W.,  \& {Allen}, S.~W. 2007, \mnras, 379, 209
\bibitem{} Schmidt, R.~W.,  \& {Allen}, S.~W. 2007, MNRAS 379, 209
\bibitem{} Sharma, S., Steinmetz, M., 2005, ApJ 628, 21
\bibitem{} Schombert, J.~M. 1988, ApJ, 328, 475
\bibitem{} Sikivie P., Tkachev I. I., Wang Y., 1997, Phys. Rev. D, 56, 1863
\bibitem{} Simon J. D., Bolatto A. D., Leroy A., Blitz L., 2003, ApJ, 596, 957 
\bibitem{} Simon J. D., Bolatto A. D., Leroy A., Blitz L., Gates E. L., 2005, ApJ, 621, 757
\bibitem{} Smith et al. 2001, ApJ 552, 493
\bibitem{} Smith, G. P., Kneib, J. P., Smail, I., Mazzotta, P., Ebeling, H., \& Czoske, O. 2005, MNRAS, 359, 417
\bibitem{} Smith, G. P., Kneib, J., Ebeling, H., Czoske, O., \& Smail, I. 2001, ApJ, 552, 493
\bibitem{} Span\'o M., Marcelin M., Amram P., Carignan C., Epinat B., Hernandez O., 2008, MNRAS, 383, 297
\bibitem{} Spekkens K., Giovanelli R., Haynes M. P., 2005, AJ, 129, 2119
\bibitem{} Spergel, D. N.; Verde, L.; Peiris, H. V.; Komatsu, E.; Nolta, M. R.; Bennett, C. L.; Halpern, M.; Hinshaw, G.; Jarosik, N.; Kogut, A.; 2003, ApJS 148, 175
\bibitem{} Stadel J., Potter D., Moore B., Diemand J., Madau P., Zemp M., Kuhlen M., Quilis V., 2009, MNRAS, 398, 21
\bibitem{} Stadel, V. J., \& Quinn, T. 2003, MNRAS, 338, 14
\bibitem{} Subramanian, K.,  Cen, R.,  and Ostriker, J.P., 2000, ApJ 538, 528 
\bibitem{} Swaters, R. A., van Albada, T. S., van der Hulst, J. M., \& Sancisi, R. 2002, A\&A, 390, 829
\bibitem{} Swaters R. A., Madore B. F., van den Bosch F. C., Balcells M., 2003a, ApJ, 583, 732
\bibitem{} Swaters R. A., Verheijen M. A.W., BershadyM. A., Andersen D. R., 2003b, ApJ, 587, L19
\bibitem{} Tonini, C., Lapi, A., and Salucci, P., 2006, ApJ 649, 591-598
\bibitem{} Treu, T., \& Koopmans, L. V. E. 2002, ApJ, 575, 87
\bibitem{} Tyson, J. A., Kochanski, G . P., \& dell'Antonio, I. P., 1998, ApJ 498, L107 
\bibitem{} Umetsu, K.,  \& {Broadhurst}, T. 2008, \apj, 684, 177
\bibitem{} Umetsu, K., et~al. 2009, \apj, 694, 1643
\bibitem{} van den Bosch F. C., Robertson B. E., Dalcanton J. J., de Blok W. J. G., 2000, AJ, 119, 1579
\bibitem{} Weinberg, M. D., \& Katz, N. 2002, ApJ, 580, 627
\bibitem{} Whiley, I.~M., Aragon-Salamanca, A., {De Lucia}, G., {et~al.} 2008, MNRAS, 387, 1253
\bibitem{} White, M., 2001, A\&A Volume 367, 27-32
\bibitem{} Williams L. L. R., Babul A., Dalcanton J. J., 2004, ApJ, 604, 18
\bibitem{} Yang, X., Mo, H.~J., \& {van den Bosch}, F.~C. 2008, ApJ, 676, 248
\bibitem{} Zappacosta, L., {Buote}, D.~A., {Gastaldello}, F., {Humphrey}, P.~J., {Bullock}, J., {Brighenti}, F.,  \& {Mathews}, W. 2006, \apj, 650, 777
%\bibitem{} Zappacosta, L., Buote, D. A., Gastaldello, F., Humphrey, P. J., Bullock, J., Brighenti, F., \& Mathews, W. 2006, ArXiv Astrophysics e-prints
\bibitem{} Zitrin, A.,  et al. 2011, MNRAS DOI: 10.1111/j.1365-2966.2011.20155.x
\end{thebibliography}

\label{lastpage}

{}

\end{document}